\documentclass[12pt,journal,draftclsnofoot,onecolumn,letterpaper]{IEEEtran} 

\usepackage{amssymb}
\usepackage{amsmath}
\usepackage{amsmath,bm}
\usepackage{amsthm}
\usepackage{graphicx}
\usepackage{subfigure}
\usepackage{cite}
\usepackage{enumerate}
\usepackage{algorithm}
\usepackage{algorithmicx}  
\usepackage{algpseudocode} 
\usepackage{color, soul}
\allowdisplaybreaks[4]
\begin{document}

\newtheorem{definition}{\bf ~~Definition}
\newtheorem{observation}{\bf ~~Observation}
\newtheorem{theorem}{\bf ~~Theorem}
\newtheorem{proposition}{\bf ~~Proposition}
\newtheorem{remark}{\bf ~~Remark}

\renewcommand{\algorithmicrequire}{\textbf{Input:}} 
\renewcommand{\algorithmicensure}{\textbf{Output:}} 

\title{IoT-U: Cellular Internet-of-Things Networks over Unlicensed Spectrum}
\author{
\IEEEauthorblockN{
{Hongliang Zhang}, \IEEEmembership{Student Member, IEEE},
{Boya Di}, \IEEEmembership{Student Member, IEEE},\\
{Kaigui Bian}, \IEEEmembership{Member, IEEE},
{and Lingyang Song}, \IEEEmembership{Fellow, IEEE}}\\

\vspace{-0.5cm}

\thanks{The authors are with National Engineering Laboratory for Big Data Analysis and Applications, School of Electronics Engineering and Computer Science, Peking University, Beijing, China (email: hongliang.zhang@pku.edu.cn; diboya@pku.edu.cn; bkg@pku.edu.cn; lingyang.song@pku.edu.cn).}}

\maketitle

\vspace{-0.5cm}

\begin{abstract}
In this paper, we consider an uplink cellular Internet-of-Things~(IoT) network, where a cellular user~(CU) can serve as the mobile data aggregator for a cluster of IoT devices. To be specific, the IoT devices can either transmit the sensory data to the base station~(BS) directly by cellular communications, or first aggregate the data to a CU through Machine-to-Machine~(M2M) communications before the CU uploads the aggregated data to the BS. To support massive connections, the IoT devices can leverage the unlicensed spectrum for M2M communications, referred to as IoT Unlicensed~(IoT-U). Aiming to maximize the number of scheduled IoT devices and meanwhile associate each IoT devices with the right CU or BS with the minimum transmit power, we first introduce a single-stage formulation that captures these objectives simultaneously. To tackle the NP-hard problem efficiently, we decouple the problem into two subproblems, which are solved by successive linear programming and convex optimization techniques, respectively. Simulation results show that the proposed IoT-U scheme can support more IoT devices than that only using the licensed spectrum.

\end{abstract}


\newpage

\section{Introduction}

We are rapidly connecting machines and other physical objects to Internet-enabled networks at unprecedented rates, accelerating towards the Internet-of-things (IoT). It is envisioned that billions of IoT devices will be connected by 2020 with transformative economic potentials for operators and stakeholders, reaching trillions of dollars \cite{M-2015}. One typical feature of the IoT network is the massive connectivity, which refers that a large number of IoT devices have accessed to the existing wireless networks. This feature has unlocked a various of opportunities to have applications across a variety of vertical sectors, such as city sensing~\cite{MKM-2018}, traffic control~\cite{OTMJM-2017}, smart home~\cite{SPS-2017}, heath care \cite{UBM-2017}, and precision agriculture~\cite{CINKK-2017}. 

To support such a large number of IoT devices, the 3rd Generation Partnership Project~(3GPP) has led the research on enabling IoT communications over the cellular systems. Its recent solutions, e.g., Narrow Band IoT~(NB-IoT), provide Quality-of-Service~(QoS) guaranteed connections and introduce power saving modes to improve battery life~\cite{AXOAEMYTAD-2017}. 
However, the licensed spectrum may not be sufficient for the mass deployments of IoT devices. To support more IoT devices, one promising solution is to extend the cellular IoT communications to the unlicensed spectrum using the LTE-Unlicensed~(LTE-U) technology \cite{HYL-2017,PBHKL-2018}, especially in the hotspot areas, which we refer to as IoT Unlicensed~(IoT-U). Specifically, the characteristics of the IoT-U systems lie in that the unlicensed carriers are integrated with the licensed ones for data transmission of IoT devices by the existing carrier aggregation~(CA) technology~\cite{RMLZXL-2015}. Besides, unlike traditional Machine-to-Machine~(M2M) communications~\cite{ZMKGL-2016}, the cellular users~(CUs) in the IoT-U network can work as mobile data aggregators to further enhance the IoT connectivity, which can aggregate the data from IoT devices.

In this paper, we consider an uplink IoT-U network\footnote{A typical application for the IoT-U networks is smart home \cite{FQ-2016}. In the application, the sensory data generated by the IoT devices at home can be aggregated to a CU of the resident by M2M communications over both licensed and unlicensed spectrum. Then, the aggregated data is forwarded to the server by cellular communications for further data processing.}, where a CU can serve as the mobile data aggregator for a cluster of IoT devices. Specifically, the CU can collect sensory data from IoT devices through M2M communications, and aggregate the data to the base station~(BS) via cellular communications. To further support more IoT devices, M2M communications can work as the underlay of cellular communications and share the unlicensed spectrum with the Wi-Fi systems. Different from the long-range communication techniques in unlicensed bands for IoT networks such as LoRa~\cite{LoRa-2015}, which builds the network upon the IEEE 802.15.4 infrastructure with a mesh topology~\cite{MLAM-2016}, IoT-U system requires the assist and control from the central BS. 

However, the spectrum utilization brings new challenges to the scheduling of IoT devices in the IoT-U network. First, a suitable coexistence mechanism of the IoT-U and Wi-Fi systems is required due to the opportunistic feature of unlicensed channel access \cite{TVOMAMSRN-2013}. Second, the interference management among CUs, IoT devices and Wi-Fi users~(WUs) becomes more complicated. To tackle the first challenge, we utilize a duty cycle \cite{Qualcomm-2014} based protocol to share the unlicensed spectrum fairly. To cope with the second one, we optimize the association, scheduling, and power allocation to maximize the weighted scheduled number of IoT devices with the lowest power consumption. This problem is a mixed-integer non-linear programming (MINLP) problem, which is generally NP-hard. To solve this problem efficiently, we decouple it into two subproblems, i.e., IoT devices association and scheduling subproblem, and power allocation subproblem. For the first subproblem, we convert the non-linear constraints into linear ones and solved the transformed problem by the branch-and-bound algorithm \cite{ED-1966}. For the second subproblem, we approximate the non-convex functions into a series of convex ones by successive convex approximation~(SCA)~\cite{JJ-2009} and solve them by existing convex techniques \cite{SL-2004}.

In literature, various techniques have been discussed for the spectrum sharing in cellular networks, such as cognitive radio \cite{MBWAF-2015,MABWH-2018}, Wi-Fi offloading \cite{XYSH-2014,KJYIS-2013}, LTE-U \cite{Nokia-2014,HXWS-2015,MSMS-2017}, and LTE-Licensed Assisted Access~(LTE-LAA) \cite{BJYJ-2017,JWSS-2018}. In \cite{MBWAF-2015,MABWH-2018}, a buffer-aided cognitive M2M communication network was considered, where the M2M communications share the spectrum with the cellular network in a cognitive manner. In \cite{XYSH-2014,KJYIS-2013}, Wi-Fi access points~(APs) was deployed for cellular networks to offload the mobile data traffic to the unlicensed spectrum. Recently, LTE-U and LTE-LAA have been proposed for the coexistence of cellular networks and Wi-Fi networks by CA. LTE-U works on the duty cycle based protocol and can only be used in China, USA, South Korea and India \cite{Nokia-2014}. In \cite{HXWS-2015}, an almost blank subframe scheme was proposed to mitigate the interference in the LTE-U network. The authors in \cite{MSMS-2017} formulated a joint user association and power allocation problem for licensed and unlicensed spectrum to maximize the LTE-U system sum-rate subjected to the QoS constraints. For worldwide deployment, LTE-LAA uses the Listen-Before-Talk (LBT) and is the modified version of LTE-U \cite{BJYJ-2017}. In \cite{JWSS-2018}, the authors proposed a novel Markov chain-based analytic model to cope with the variation of the LTE-LAA frame structure. Different from the existing schemes that focused on either M2M or cellular links, our proposed IoT-U scheme considers the the network assisted data aggregation on both the licensed and unlicensed spectrum, thereby utilizing the spectrum more efficiently.

Therefore, the major contributions of this paper are summarized as follows.
\begin{enumerate}
	\item We consider an uplink IoT-U network to support more IoT devices, where the CUs can aggregate sensory data from IoT devices over both licensed and unlicensed spectrum, and utilize a feasible duty cycle based protocol to facilitate the fair unlicensed spectrum between the IoT-U and Wi-Fi systems.
	
	\item We maximize the weighted scheduled number of IoT devices in each cycle by jointly optimizing the IoT device association, scheduling, and power allocation. To solve the problem efficiently, we decouple the problem into IoT devices association and scheduling as well as power allocation subproblems, and propose an iterative algorithm to solve these subproblems. 
	
	\item Simulation results show that our proposed IoT-U scheme can achieve a better performance than the IoT scheme without unlicensed band, decoupled unlicensed scheme, and the matching based unlicensed scheme.
\end{enumerate}

The rest of the paper is organized as follows. In Section \ref{System}, we first introduce the system model for the uplink IoT-U network. The spectrum sharing mechanism for IoT-U and Wi-Fi systems is introduced in Section \ref{mechanism} together with the interference analysis. In Section \ref{problem}, we introduce the problem formulation and decouple the problem into two subproblems: IoT devices association and scheduling problem, and power allocation subproblem. Section \ref{Algorithm} presents an iterative algorithm to solve these two subproblems. In Section \ref{Discussion}, the system performance is discussed. Numerical results in Section \ref{sec:simulation} evaluate the performance of our proposed IoT-U scheme. Finally, conclusions are drawn in Section \ref{sec:conclusion}.

\section{System Model}%
\label{System}

\begin{figure}[!t]
\centering
\includegraphics[width=5.0in]{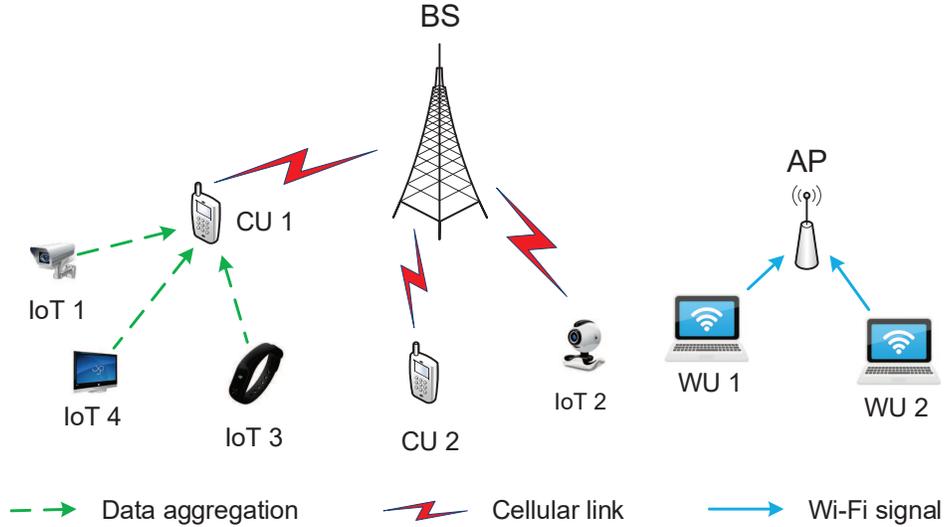}
\caption{System model for the uplink IoT-U network.}
\label{scenario}
\end{figure}

As shown in Fig.~\ref{scenario}, we consider an uplink IoT-U network consisting of one BS, $N$ CUs, denoted by $\mathcal{N} = \{1,\ldots,N\}$, and $M$ IoT devices to collect sensory data, denoted by $\mathcal{M} = \{1,\ldots,M\}$. Among these CUs, there are $Q$ idle CUs which have no data to transmit and can serve as aggregators to aggregate the data generated from IoT devices, denoted by $\mathcal{Q} = \{1,\ldots,Q\}$. To upload the sensory data efficiently, there exist two communication modes as described below:
\begin{itemize}
	\item Cellular mode: The IoT devices can transmit their sensory data to the BS directly by cellular communications.
	
	\item Aggregation mode: An idle CU can aggregate the sensory data through M2M communications\footnote{The relaying network has been investigated in 3GPP to extend the coverage of IoT devices~\cite{3GPP-2017}. The relaying CU and IoT devices belong to the same group, for example, they subscribe to the same mobile operator, and thus, the CU is willing to relay the data traffic for IoT devices.} before the CU uploads the aggregated data to the BS.
\end{itemize}
In Fig.~\ref{scenario}, for example, the collected data of IoT device 2 is sent to BS directly, while IoT devices 1, 3, and 4 form a cluster and their sensory data is sent to the BS by data aggregation to CU 1. The network owns $K$ licensed subchannels to support orthogonal frequency division multiple access~(OFDMA) transmissions, denoted by $\mathcal{K} = \{1,\ldots,K\}$. Different cellular links need to utilize orthogonal channels to avoid mutual interference.

For the Wi-Fi system, we assume that there exist $F$ active WUs marked by $\mathcal{F} = \{1, \ldots, F\}$. Besides, we assume that there are $L$ unlicensed channels to support these WUs, e.g., there are 23 channels for IEEE 802.11n in the 5GHz band. In each cluster, the M2M links among IoT devices and the CUs can utilize both licensed and unlicensed spectrum via CA. Since the bandwidth of an unlicensed channel is much wider than a licensed subchannel, to utilize the unlicensed channel more efficiently, we divide the unlicensed channel into $K^u$ subchannels with the same bandwidth as the licensed one, denoted by $\mathcal{K}^u = \{1,\ldots,K^u\}$~\cite{HYL-2017}. To make full use of the licensed band, we assume that the M2M links can share the licensed band with the cellular links. Besides, we also assume that the CUs can upload the data to the BS and aggregate the data from IoT devices simultaneously on different subchannels. 

We define $\bm{P}_{(M \times (N + 1) + N) \times T \times (K + K^u)} = \left(\bm{P}^C_{N \times T \times (K + K^u)},\bm{P}^I_{M \times (N + 1) \times T \times (K + K^u)} \right)$ as the transmit power matrix, where $\bm{P}^C_{N \times T \times (K + K^u)} = [p_{n}^{k,t}]$ implies the transmit power of the link between CU~$n$ and the BS over subchannel $k$ in subframe $t$, and $\bm{P}^I_{M \times (N + 1) \times T \times (K + K^u)} = [p_{m,n}^{k,t}]$ implies the transmit power of the link between IoT device $m$ and CU $n$ over subchannel $k$ in subframe $t$. Here, $n = 0$ if the IoT device $m$ transmits to the BS directly. We also denote the total transmit powers of IoT device $m$ and CU $n$ by $P^{iot}$ and $P^c$, respectively. The transmit power of WUs on the whole unlicensed channel is fixed as $P^w$. 
Therefore, we have
\begin{equation}
\begin{array}{ll}
&\sum\limits_{k \in \mathcal{K}} p_{m,0}^{k,t} + \sum\limits_{k \in \mathcal{K} \cup \mathcal{K}^u} \sum\limits_{n \in \mathcal{N}} p_{m,n}^{k,t} \leq P^{iot},\\
&\sum\limits_{k \in \mathcal{K}}p_{n}^{k,t} \leq  P^c.
\end{array}
\end{equation}

The free space propagation path-loss model with Rayleigh fading is adopted to model the channel gain between two nodes in the network, i.e., for the link from nodes $i$ to $j$ over subchannel $k$, the received power can be expressed as
\begin{equation}\label{Received_power}
p_{i,j}^k = p^k_i \cdot |h_{i,j}^k|^2 = p^k_i \cdot G \cdot d_{i,j}^{-\alpha} \cdot |h^k_0|^2,
\end{equation}
where $p^k_i$ represents the transmit power of user $i$ over subchannel $k$, $d_{i,j}$ is the distance between nodes $i$ and $j$, $\alpha$ is the pathloss exponent, $G$ is the constant power gains factor introduced by amplifier and antenna, and $h^k_0 \sim \mathcal{CN}(0,1)$ is a complex Gaussian variable representing Rayleigh fading. In addition, we assume that the thermal noise at each user satisfies independent Gaussian distribution with zero mean and the same variance $\sigma ^2$.

\section{Spectrum Sharing Mechanism And Interference Analysis}
\label{mechanism}

In this section, we first introduce the spectrum sharing mechanism for IoT-U and Wi-Fi systems. Then, the interference issues over licensed and unlicensed spectrum are discussed, respectively.

\subsection{Unlicensed Spectrum Sharing Mechanism}
\begin{figure}[!t]
	\centering
	\includegraphics[width=4.5in]{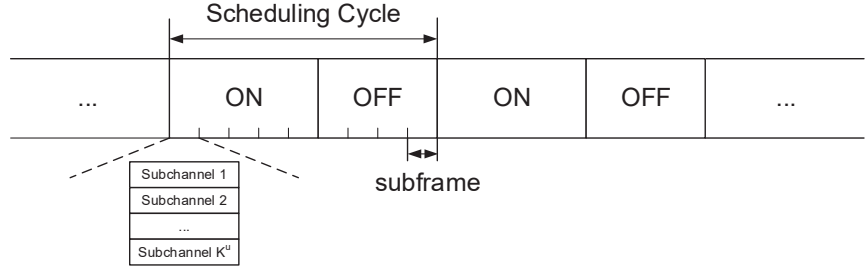}
	\caption{Duty cycle based spectrum sharing mechanism for IoT devices in the unlicensed spectrum.}
	\label{protocol}
\end{figure}

In the unlicensed spectrum, the IoT devices are allowed to share channels with WUs. To share the unlicensed spectrum fairly, a duty cycle method is adopted to support the coexistence of the IoT-U and Wi-Fi systems. Specifically, as shown in Fig.~\ref{protocol}, the unlicensed band is slotted to subframes whose time length is the same as that in the cellular system for the sake of synchronization. To schedule the IoT devices efficiently, we define the concept of scheduling cycle containing $T$ subframes, denoted by $\mathcal{T} = \{1,\ldots,T\}$, in which each IoT device will be scheduled at most once. 

At the beginning of a scheduling cycle, the BS will select a channel with the least interference first according to the channel measurement by the IoT devices. The IoT devices will scan the unlicensed spectrum and identify a channel with the least interference from the set of available unlicensed channels for the uplink transmission. Specifically, the IoT devices will perform energy detection, and measure the interference level. If the interference is sensed less than the predefined threshold for a sensing duration, the channel will be regarded as idle for this IoT device. Then, these IoT devices will inform the BS whether they collide with WUs according to the measured result. If in the operating channel, the number of interfered users is larger than a given threshold, and there is another available channel with the less interference, the transmission will be switched to the new channel.

The scheduling cycle consists of two transmission modes, i.e., ON and OFF modes. In the ON mode, the IoT devices are enabled to transmit on unlicensed spectrum. The time period for the ON mode contains $T_{ON}$ subframes, denoted by $\mathcal{T}_{ON} = \{1,\ldots,T_{ON}\}$, and each of them is divided into $K^u$ subchannels to support the concurrent transmissions for different IoT devices. However, in the OFF mode, the IoT devices need to be silent in the unlicensed channel, which is handed over to the Wi-Fi system. When the reserved period for the Wi-Fi system expires, the IoT devices are activated in the unlicensed subchannels. In this way, the IoT-U system coexists with the Wi-Fi system in a time division multiplexing~(TDM) manner. The duty cycle percentage $T_{ON}/T$ captures the effect of IoT-U transmission on the Wi-Fi system, and should be carefully designed to provide fair unlicensed resource sharing. 

To evaluate the signaling overhead for the spectrum sharing mechanism, we assume that $\alpha$ messages are required to inform the BS of the channel state sensed by an IoT device, $\beta$ messages are required for an IoT device/CU to report its location and subchannel estimation results over a subchannel, and $x$ messages are necessary for the BS to notify a user about the allocated subchannels. At the beginning of a scheduling cycle, each IoT device needs to report the sensing result over each channel, and thus, at most $\alpha ML$ messages are required. And before each subframe, each IoT device/CU needs to report their locations and the subchannel estimation results for subchannel allocation, which requires $\beta(M + N)(T - T_{ON})K + \beta T_{ON}(M(K + K^u) + NK)$ messages for a cycle. Then, the BS will perform resource allocation process with extra information, and notify each user by sending at most $x(M + N)T$ messages. Note that in one duty cycle, each IoT device only performs one energy detection over one channel. Thus, the signaling cost is under a tolerable level. In addition, the signaling cost of resource allocation is positively proportional to the number of IoT devices and CUs as well as the number of available unlicensed subchannels. When the number of IoT devices is numerous, we can also reduce the signaling overhead by constraining the number of available unlicensed subchannels for each IoT device.

\subsection{Interference Analysis}

In this part, we will discuss the interference over licensed and unlicensed spectrum.

\subsubsection{Transmission Model over Licensed Band}

In licensed spectrum, cellular and M2M links are able to occupy multiple subchannels for transmission, and a subchannel can be allocated to one cellular link. To guarantee reliable transmission of the control signaling, a M2M link is required to hold at least one licensed subchannel~\cite{RMLZXL-2015}. Besides, M2M links can work as an underlay of the cellular links. In other words, M2M links can utilize the licensed subchannels concurrently occupied by some cellular links.

Under these assumptions, the cellular links can only receive the co-channel interference from the underlaid M2M links, while the interference received by M2M links might be from cellular links and other co-channel M2M links. Define the set of IoT devices which upload sensory data by cellular communications as $\mathcal{S}^C$ and the set of IoT devices which transmit sensory data by M2M communications as $\mathcal{S}^M$, respectively. In addition, we define the IoT device association and subchannel allocation matrix for licensed band $\bm{A}_{(M \times (Q + 1) + N) \times T \times K} = \left(\bm{\Phi}_{N \times T \times K},\bm{\Theta}_{M \times (Q + 1) \times T \times K} \right)$, where $\bm{\Phi}_{N \times T \times K} = [\phi^k_{n,t}]$, and $\bm{\Theta}_{M \times (Q + 1) \times T \times K} = [\theta^{k,t}_{m,n}]$ stand for the subchannel allocation matrices for CUs and IoT devices, respectively. The values of $\phi^{k,t}_{n}$ and $\theta^{k,t}_{m,n}$ are defined as
\begin{equation}\label{phi}
\phi^{k,t}_{n}=
\left\{ \begin{gathered}
1,~~\mbox{when licensed subchannel}~k~\mbox{is allocated to CU}~n~\mbox{in subframe}~t,\hfill\\
0,~~\mbox{otherwise},\hfill\\
\end{gathered} \right.
\end{equation}
and
\begin{equation}\label{theta}
\theta^{k,t}_{m,n}=
\left\{ \begin{gathered}
1,~~\mbox{when licensed subchannel}~k~\mbox{is allocated to link}~m-n~\mbox{in subframe}~t,\hfill\\
0,~~\mbox{otherwise}.\hfill\\
\end{gathered} \right.
\end{equation}
 
Using the aforementioned notations, the SINR at the receiver of BS from CU $n$ over licensed subchannel $k$ in subframe $t$ can be given by
\begin{equation}\label{SINR_CU}
\gamma_{n}^{k,t}  = \frac{{p_{n}^{k,t}\phi_{n}^{k,t} |h^{k,t}_{n}|^2 }}{{\sigma ^2  + \sum\limits_{m \in \mathcal{S}^M} \sum\limits_{\begin{subarray}{l}
			n' \in \mathcal{Q},\\
			n' \ne n
			\end{subarray}} {p_{m,{n'}}^{k,t} \theta_{m,{n'}}^{k,t} |h^{k,t}_{m,0}|^2 } }},
\end{equation}
where $h^{k,t}_{n}$ and $h^{k,t}_{m,0}$ represent the channel gains from CU $n$ and IoT device $m$ to the BS over subchannel $k$ in subframe $t$, respectively. Likewise, the SINR at the receiver of the BS from IoT device $m \in \mathcal{S}^C$ over licensed subchannel $k$ in subframe $t$ can be expressed as
\begin{equation}\label{SINR_IoT_cellular}
\gamma_{m,0}^{k,t}  = \frac{{ p_{m,0}^{k,t} \theta_{m,0}^{k,t} |h^k_{m,0}|^2}}{{\sigma ^2  + \sum\limits_{m' \in \mathcal{S}^M} \sum\limits_{n \in \mathcal{Q}} {p_{m',n}^{k,t} \theta_{{m'},n}^{k,t} |h^{k,t}_{{m'},0}|^2}}}.
\end{equation}
In addition, the SINR at the associated CU $n \in \mathcal{Q}$ from IoT $m \in \mathcal{S}^M$ over licensed subchannel $k$ in subframe $t$ can be given by
\begin{equation}\label{SINR_IoT_M2M}
\gamma_{m,n}^{k,t} = \frac{p_{m,n}^{k,t} \theta_{m,n}^{k,t}|h^{k,t}_{m,n}|^2}{\sigma ^2  + \sum\limits_{\begin{subarray}{l}
		n' \in \mathcal{N},\\
		n' \ne n
		\end{subarray}}{p_{{n'}}^{k,t} \phi_{{n'}}^{k,t}|h^{k,t}_{{n'},n}|^2} + \sum\limits_{\begin{subarray}{l}
		m' \in \mathcal{M},\\
		m' \ne m
		\end{subarray}} \sum\limits_{\begin{subarray}{l}
		n' \in \mathcal{Q} \cup \{0\},\\
		n' \ne n
		\end{subarray}} {P_{{m'},{n'}}^{k,t} \theta_{{m'},{n'}}^{k,t} |h^{k,t}_{{m'},{n}}|^2}},
\end{equation}
where $h^{k,t}_{m,n}$ and $h^{k,t}_{{n'},n}$ represent the channel gains from IoT device $m$ and CU $n'$ to CU $n$ over subchannel $k$ in subframe $t$, respectively. Therefore, the achievable rates of the CU $n$ - BS, IoT device $m$ - BS or associated CU $n$ links over subchannel $k$ in subframe $t$ can be obtained by
\begin{equation}
R_{n}^{k,t} = \log_2(1 + \gamma_{n}^{k,t}),~~R_{m,n}^{k,t} = \log_2(1 + \gamma_{m,n}^{k,t}),
\end{equation}
respectively.

\subsubsection{Transmission Model over Unlicensed Band}

In the unlicensed spectrum, only IoT devices which set up M2M communications with CUs are allowed to transmit. Therefore, the interference for a M2M link only comes from other M2M links utilizing the same unlicensed subchannel. Define the subchannel allocation matrix for unlicensed band $\bm{B}_{M \times Q \times T \times K^u} = [\beta^{k,t}_{m,n}]$, where
\begin{equation}\label{beta}
\beta^{k,t}_{m,n}=
\left\{ \begin{gathered}
1,~~\mbox{when unlicensed subchannel}~k~\mbox{is allocated to link}~m-n~\mbox{in subframe}~t,\hfill\\
0,~~\mbox{otherwise},\hfill\\
\end{gathered} \right.
\end{equation}
The SINR at the associated CU from IoT device $m$ over unlicensed subchannel $k$ in subframe $t$ can be written as
\begin{equation}\label{SINR_M2M_Unlicensed}
\gamma_{m,n}^{k,t,u}  = \frac{p_{m,n}^{k,t} \beta_{m,n}^{k,t} |h^{k,t,u}_{m,n}|^2}{\sigma ^2  + \sum\limits_{\begin{subarray}{l}
		m' \in \mathcal{M},\\
		m' \ne m
		\end{subarray}} \sum\limits_{\begin{subarray}{l}
		n' \in \mathcal{Q},\\
		n' \ne n
		\end{subarray}}{p_{{m'},{n'}}^{k,t} \beta^{k,t}_{{m'},{n'}} |h^{k,t,u}_{{m'},{n}}|^2} + I^{k,t}_{m,n} },
\end{equation}
where $h^{k,t,u}_{m,n}$ is channel gain from IoT device $m$ to CU $n$ over unlicensed subchannel $k$ and $I^k_m$ is the total interference from Wi-Fi system to the associated CU of IoT device $m$. Here, the interference can be calculated as 
\begin{equation}
I^{k,t}_{m,n} = \sum\limits_{f \in \mathcal{F}} {\frac{{P^w }}{{K^u }}} |h^{k,t}_{f,n}|^2, 
\end{equation}
where $h^{k,t}_{f,n}$ is the channel gain from the active WU $f$ to CU $n$ over subchannel $k$ in subframe $t$. Thus, the data rate of IoT device $m$ over unlicensed subchannel $k$ in subframe $t$ is given by
\begin{equation}\label{rate_unlicensed}
R_{m,n}^{k,t,u} = \log_2(1 + \gamma_{m,n}^{k,t,u}).
\end{equation}

\section{Weighted Scheduled Number Maximization of IoT Devices}%
\label{problem}

In this section, we aim to maximize the weighted scheduled number of IoT devices in each cycle by jointly optimizing the IoT device association, scheduling, and power allocation.

\subsection{Constraints for Clusters}

To guarantee the performance of the IoT-U system, several constraints in terms of the data rate, subchannel allocation, and scheduling are illustrated below.

In this paper, we assume that the CUs can aggregate data from the IoT devices and upload the data to the BS simultaneously. To avoid the mutual interference, a cellular link and the M2M links associated to the same CU need to utilize different licensed subchannels, i.e.,
\begin{equation}
\phi^{k,t}_{n} + \sum\limits_{m \in \mathcal{M}} \theta^{k,t}_{m,n} \leq 1, \forall n \in \mathcal{Q}, k \in \mathcal{K}, t \in \mathcal{T}.\label{subchannel1}
\end{equation}
Besides, following the OFDMA manner, the IoT devices associated with the same CU cannot utilize the same unlicensed subchannel, and thus, we have 
\begin{equation}
\sum\limits_{m \in \mathcal{M}} \beta^{k,t}_{m,n} \leq 1, \forall n \in \mathcal{Q}, k \in \mathcal{K}^u, t \in \mathcal{T}.
\end{equation}
Similarly, we assume that a licensed subchannel can be allocated to at most one cellular link. Therefore, the subchannel allocation matrix for the licensed spectrum satisfies
\begin{equation}
\sum\limits_{n \in \mathcal{N}}\phi^{k,t}_{n} + \sum\limits_{m \in \mathcal{M}} \theta_{m,0}^{k,t} \leq 1, \forall k \in \mathcal{K}, t \in \mathcal{T}.\label{subchannel2}
\end{equation}

Define a association matrix $\bm{F}_{M \times (Q + 1) \times T} = [f_{m,n}^t]$ to indicate whether IoT device $m$ associates with CU $n$ in subframe $t$, where
\begin{equation}\label{F}
f_{m,n}^t=
\left\{ \begin{gathered}
1,~~\mbox{when IoT device}~m~\mbox{associates with CU}~n~\mbox{in subframe}~t,\hfill\\
0,~~\mbox{otherwise},\hfill\\
\end{gathered} \right.
\end{equation}
Thus, we have
\begin{equation}\label{association2}
1 - f_{m,n}^t =  \prod\limits_{k \in \mathcal{K}} (1 - \theta_{m,n}^{k,t}), \forall m \in \mathcal{M}, \forall n \in \mathcal{Q} \cup \{0\}, \forall t \in \mathcal{T} .
\end{equation}

Likewise, define a scheduling matrix $\bm{C}_{M \times T} = [c_{m,t}]$ to indicate whether IoT device $m$ is scheduled in subframe $t$, where
\begin{equation}\label{c}
c_m^t=
\left\{ \begin{gathered}
1,~~\mbox{when IoT device}~m~\mbox{is scheduled in subframe}~t,\hfill\\
0,~~\mbox{otherwise},\hfill\\
\end{gathered} \right.
\end{equation}
Note that a IoT device is assigned at least one licensed subchannel if it is scheduled. Therefore, we have
\begin{equation}\label{relation}
1 - c_{m,t} = \prod\limits_{n \in \mathcal{Q} \cup \{0\}} (1 - f_{m,n}^{t}), \forall m \in \mathcal{M}, \forall t \in \mathcal{T}.
\end{equation}
According to the requirement of CA, each M2M link needs to occupy at least one licensed subchannel for control signaling, and thus, we have
\begin{equation}
\sum\limits_{k \in \mathcal{K}}\sum\limits_{n \in \mathcal{Q}}\theta^{k,t}_{m,n} \geq c_{m,t}, \forall m \in \mathcal{S}^M, t \in \mathcal{T}_{ON}.
\end{equation}
Since an IoT device can be scheduled at most once in a cycle, we also have a scheduling constraint:
\begin{equation}\label{scheduling}
\sum\limits_{t \in \mathcal{T}}c_{m,t} \leq 1, \forall m \in \mathcal{M}.
\end{equation}
In addition, a scheduled IoT device can associate with at most one CU or BS, therefore, the association matrix needs to satisfy
\begin{equation}\label{association}
\sum\limits_{n \in \mathcal{Q} \cup \{0\}} f_{m,n}^{t} \leq c_{m,t}, \forall m \in \mathcal{M},  \forall t \in \mathcal{T}.
\end{equation}

Moreover, for each IoT device, the achievable data rate needs to exceed the generated data in order to upload the sensory data in a subframe. Denote the generated data of IoT device $m$ by $d_m$, and therefore, for IoT devices in the cellular mode, we have
\begin{equation}
\sum\limits_{k \in \mathcal{K}} R_{m,0}^{k,t} \geq d_m c_{m,t}, \forall m \in \mathcal{S}^C, t \in \mathcal{T},\label{rate1}
\end{equation}
and for IoT devices in the aggregation mode, the following equations should be satisfied, which correspond to the subframes in the ON and OFF modes, respectively.
\begin{align}
&\sum\limits_{k \in \mathcal{K}} \sum\limits_{n \in \mathcal{Q}} R_{m,n}^{k,t} + \sum\limits_{k \in \mathcal{K}^u} \sum\limits_{n \in \mathcal{Q}} R_{m,n}^{k,t,u} \geq d_m c_{m,t}, \forall m \in \mathcal{S}^M, t \in \mathcal{T}_{ON},\\
&\sum\limits_{k \in \mathcal{K}} \sum\limits_{n \in \mathcal{Q}} R_{m,n}^{k,t} \geq d_m c_{m,t}, \forall m \in \mathcal{S}^M, t \in \mathcal{T} \backslash \mathcal{T}_{ON}.
\end{align}
 In addition, to transmit the aggregated data to the BS successfully, the achievable data rate of the CU should be larger than the aggregated data in both the ON and OFF modes, i.e., 
\begin{align}
&\sum\limits_{k \in \mathcal{K}} R_{n}^{k,t} \geq \sum\limits_{m \in \mathcal{S}^M} \sum\limits_{k \in \mathcal{K} \cup \mathcal{K}^u} R_{m,n}^{k,t}, \forall n \in \mathcal{Q}, t \in \mathcal{T}_{ON},\\
&\sum\limits_{k \in \mathcal{K}} R_{n}^{k,t} \geq \sum\limits_{m \in \mathcal{S}^M} \sum\limits_{k \in \mathcal{K}} R_{m,n}^{k,t}, \forall n \in \mathcal{Q}, t \in \mathcal{T} \backslash \mathcal{T}_{ON}.
\end{align}
Define the minimum rate requirement for an active CU as $R_{min}$. To guarantee the QoS of active CUs, we have the constraint below:
\begin{equation}
\sum\limits_{k \in \mathcal{K}}R_{n}^{k,t} \geq R_{min}, \forall n \in \mathcal{N} \backslash \mathcal{Q}, t \in \mathcal{T}.  \label{rate2}
\end{equation}

\subsection{Problem Formulation}

To support the massive number of IoT devices, we aim to schedule more IoT devices in one cycle. However, those IoT devices with better channel conditions will be scheduled more frequently because they need fewer subchannels, and thus, more IoT devices can be scheduled. In this paper, we consider the user fairness in terms of scheduled opportunity, and introduce a weight factor $w_m$ for IoT device $m$ to adjust their scheduling priority. Based on the proportional fairness scheduler~\cite{BLY-2016}, the value of $w_m$ is defined to be inversely proportional to the average scheduled times of IoT device $m$ in the previous cycles. Therefore, the system performance can be evaluated by the weighted number of scheduled IoT devices in a cycle. Then, the scheduling problem can be formulated as
\begin{subequations}
\begin{align}
\text{P1}:\max\limits_{\bm{A},\bm{B},\bm{C},\bm{F},\bm{P}} & \sum\limits_{m \in \mathcal{M}}w_m\sum\limits_{t \in \mathcal{T}}c_{m,t}, \label{Objective} \\
s.t.~~&\sum\limits_{k \in \mathcal{K}} p_{m,0}^{k,t} + \sum\limits_{k \in \mathcal{K} \cup \mathcal{K}^u} \sum \limits_{n' \in \mathcal{Q}} p_{m,{n'}}^{k,t} \leq P^{iot}, \sum\limits_{k \in \mathcal{K} }p_{n}^{k,t} \leq P^c, \forall m \in \mathcal{M}, n \in \mathcal{N}, t \in \mathcal{T}_{ON}, \label{constraint1}\\
&\sum\limits_{k \in \mathcal{K}} \sum \limits_{n' \in \mathcal{Q} \cup \{0\}} p_{m,{n'}}^{k,t} \leq P^{iot}, \sum\limits_{k \in \mathcal{K}}p_{n}^{k,t} \leq P^c, \forall m \in \mathcal{M}, n \in \mathcal{N}, t \in \mathcal{T} \backslash \mathcal{T}_{ON}, \label{constraint2}\\
&\phi^{k,t}_{n} + \sum\limits_{m \in \mathcal{M}} \theta^{k,t}_{m,n} \leq 1, \forall n \in \mathcal{Q}, k \in \mathcal{K}, t \in \mathcal{T},\label{constraint3}\\
&\sum\limits_{m \in \mathcal{M}} \beta^{k,t}_{m,n} \leq 1, \forall n \in \mathcal{Q}, k \in \mathcal{K}^u, t \in \mathcal{T},\label{constraint4}\\
&\sum\limits_{n \in \mathcal{N}}\phi^{k,t}_{n} + \sum\limits_{m \in \mathcal{M}} \theta_{m,0}^{k,t} \leq 1, \forall k \in \mathcal{K}, t \in \mathcal{T},\label{constraint5}\\
&\sum\limits_{k \in \mathcal{K}}\sum\limits_{n \in \mathcal{Q}}\theta^{k,t}_{m,n} \geq c_{m,t}, \forall m \in \mathcal{S}^M, t \in \mathcal{T},\label{constraint6}\\
&\sum\limits_{t \in \mathcal{T}}c_{m,t} \leq 1, \forall m \in \mathcal{M},\label{constraint7}\\
&\sum\limits_{n \in \mathcal{Q} \cup \{0\}} f_{m,n}^{t} \leq c_{m,t}, \forall m \in \mathcal{M}, \forall k \in \mathcal{K} \cup \mathcal{K}^u, \forall t \in \mathcal{T},\label{constraint8}\\
&1 - c_{m,t} = \prod\limits_{n \in \mathcal{Q} \cup \{0\}} (1 - f_{m,n}^{t}), \forall m \in \mathcal{M}, \forall t \in \mathcal{T}, \label{constraint9}\\
&1 - f_{m,n}^t =  \prod\limits_{k \in \mathcal{K}} (1 - \theta_{m,n}^{k,t}), \forall m \in \mathcal{M}, \forall n \in \mathcal{Q} \cup \{0\}, \forall t \in \mathcal{T},\label{constraint10}\\ 
& \sum\limits_{k \in \mathcal{K}} R_{m,0}^{k,t} \geq d_m c_{m,t}, \forall m \in \mathcal{S}^C, t \in \mathcal{T},\label{constraint11}\\
&\sum\limits_{k \in \mathcal{K}} \sum\limits_{n \in \mathcal{Q}} R_{m,n}^{k,t} + \sum\limits_{k \in \mathcal{K}^u} \sum\limits_{n \in \mathcal{Q}} R_{m,n}^{k,t,u} \geq d_m c_{m,t}, \forall m \in \mathcal{S}^M, t \in \mathcal{T}_{ON},\label{constraint12}\\
&\sum\limits_{k \in \mathcal{K}} \sum\limits_{n \in \mathcal{Q}} R_{m,n}^{k,t} \geq d_m c_{m,t}, \forall m \in \mathcal{S}^M, t \in \mathcal{T} \backslash \mathcal{T}_{ON},\label{constraint13}\\
&\sum\limits_{k \in \mathcal{K}} R_{n}^{k,t} \geq \sum\limits_{m \in \mathcal{S}^M} \sum\limits_{k \in \mathcal{K} \cup \mathcal{K}^u} R_{m,n}^{k,t}, \forall n \in \mathcal{Q}, t \in \mathcal{T}_{ON},\label{constraint14}\\
&\sum\limits_{k \in \mathcal{K}} R_{n}^{k,t} \geq \sum\limits_{m \in \mathcal{S}^M} \sum\limits_{k \in \mathcal{K}} R_{m,n}^{k,t}, \forall n \in \mathcal{Q}, t \in \mathcal{T} \backslash \mathcal{T}_{ON},\label{constraint15}\\
&\sum\limits_{k \in \mathcal{K}}R_{n}^{k,t} \geq R_{min}, \forall n \in \mathcal{N} \backslash \mathcal{Q}, t \in \mathcal{T},\label{constraint16}\\
&\phi_{n}^{k,t},\theta_{m,n}^{k,t},\beta_{m,n}^{k,t},f_{m,n}^{t},c_{m,t} \in \{0,1\}, p_n^{k,t} \geq 0, p_{m,n}^{k,t} \geq 0.\label{constraint17}
\end{align}
\end{subequations}
Constraints (\ref{constraint1}) and (\ref{constraint2}) are the power allocation constraints for the ON and OFF modes, respectively. Constraints (\ref{constraint3})- (\ref{constraint5}) are the subchannel constraints corresponding to (\ref{subchannel1})-(\ref{subchannel2}). According to the property of CA, each M2M link needs to occupy at least one licensed subchannel for control signals, and thus, constraint (\ref{constraint6}) needs to be satisfied. Constraints (\ref{constraint7})-(\ref{constraint10}) are scheduling and association constraints corresponding to (\ref{association2}), (\ref{relation}), (\ref{scheduling}), and (\ref{association}), respectively. Constraints (\ref{constraint11})-(\ref{constraint16}) are rate constraints corresponding to (\ref{rate1})-(\ref{rate2}). 

In general, the solution to problem (P1) is not unique, since different IoT device association strategies may result in the same number of scheduled IoT devices. Denote the optimal solution set of problem (P1) as $\Omega$, where
\begin{equation}
\Omega = \{(\bm{A},\bm{B},\bm{C},\bm{F},\bm{P})|\forall(\bm{A},\bm{B},\bm{C},\bm{F},\bm{P})~\text{is a solution to problem (P1)}\}.
\end{equation}
To reduce power consumption of the IoT system, it is desirable to find one solution in $\Omega$ that consumes the minimum amount of power for the IoT devices and CUs. This can be achieved by solving the following power control problem:
\begin{equation}
\text{P2:} \min\limits_{(\bm{A},\bm{B},\bm{C},\bm{F},\bm{P}) \in \Omega} \sum\limits_{m \in \mathcal{M}} \sum\limits_{t \in \mathcal{T}} \sum\limits_{k \in \mathcal{K} \cup \mathcal{K}^u} \sum\limits_{n \in \mathcal{Q} \cup \{0\}} p_{m,n}^{k,t} +\sum\limits_{n \in \mathcal{N}} \sum\limits_{t \in \mathcal{T}} \sum\limits_{k \in \mathcal{K}} p_{n}^{k,t}.
\end{equation}
Through solving problems (P1) and (P2) sequentially, we can achieve the maximum system utility with the minimum transmission power. Problem~(P1) is a mixed-integer non-convex problem due to the products of resource allocation variables in the constraints. In this paper, we adopt the single-stage formulation technique \cite{LYYJ-2013} to combine problems (P1) and (P2) into a single-stage optimization problem, which is easier to tackle with.

The utility of the single-stage optimization problem can expressed as
\begin{equation}
U = \sum\limits_{m \in \mathcal{M}} \sum\limits_{t \in \mathcal{T}} \sum\limits_{k \in \mathcal{K} \cup \mathcal{K}^u} \sum\limits_{n \in \mathcal{Q} \cup \{0\}} \hspace{-3mm}\epsilon p_{m,n}^{k,t} +\sum\limits_{n \in \mathcal{N}} \sum\limits_{t \in \mathcal{T}} \sum\limits_{k \in \mathcal{K}} \epsilon p_{n}^{k,t} + (1-\epsilon) \sum\limits_{m \in \mathcal{M}}\lambda_m\sum\limits_{t \in \mathcal{T}}(1 - c_{m,t}),
\end{equation}
where $\lambda_m$'s are positive integers obtained by scaling up $w_m$ by the same constant $\rho$, i.e.,
\begin{equation}
\rho \triangleq \frac{\lambda_m}{w_m} , \forall m \in \mathcal{M}.
\end{equation}
Thus, the single-stage problem be formulated as
\begin{equation}
\begin{array}{ll}
	\text{P3}:&\min\limits_{\bm{A},\bm{B},\bm{C},\bm{F},\bm{P}} U,  \\
	s.t. &\text{(\ref{constraint1})-(\ref{constraint17})},
\end{array}
\end{equation}

The following theorem shows that solving problem (P3) is equivalent to solving problems (P1)
and (P2) sequentially, as long as $\epsilon$ is a constant satisfying  
\begin{equation}\label{ep}
0 < \epsilon < \frac{1}{MP^{iot} + NP^{c} + 1}. 
\end{equation}

\begin{theorem}\label{equilvant}
	Given $\epsilon$ as in (\ref{ep}), the solution to problem (P3) is equivalent to solving problems (P1) and (P2) sequentially.
\end{theorem}
\begin{IEEEproof}
	See Appendix \ref{proof_equilvant}.
\end{IEEEproof}

\subsection{Problem Decomposition}
Since problem (P3) is an MINLP problem, it is NP-hard. Besides, the IoT devices association, scheduling, and power allocation variables are nested. To solve problem (P3) efficiently, we decompose it into two subproblems: IoT device association and scheduling subproblem, and power allocation subproblem. 

\subsubsection{IoT Device Association and Scheduling Subproblem}
Given the power allocation $\bm{P}$, problem (P3) can be written by
\begin{equation}
	\begin{array}{ll}
	\text{SP1}:&\min\limits_{\bm{A}, \bm{B}, \bm{C}, \bm{F}} (1-\epsilon) \sum\limits_{m \in \mathcal{M}}\lambda_m\sum\limits_{t \in \mathcal{T}}(1 - c_{m,t}),  \\
	s.t.~~& \text{(\ref{constraint3})-(\ref{constraint16})},\\
	& \phi^{k,t}_{n},\theta^{k,t}_{m,n},\beta^{k,t}_{m,n}, f_{m,n}^t, c_{m,t} \in \{0,1\}.
	\end{array}
\end{equation}

Motivated by the decomposition technique in \cite{MSAJ-2007}, define $\delta_m$ as the Lagrangian multiplier corresponding to constraint (\ref{constraint7}). Thus, the Lagrangian dual problem can be written by
\begin{equation}
	\begin{array}{ll}
	\text{SP2}:\max\limits_{\delta_m \geq 0} &\sum\limits_{t \in \mathcal{T}} \min\limits_{\bm{A}, \bm{B}, \bm{C}, \bm{F}} \left((1-\epsilon) \sum\limits_{m \in \mathcal{M}}\lambda_m (1 - c_{m,t}) - \sum \limits_{m \in \mathcal{M}} \delta_m (c_{m,t} - 1/T)\right),  \\
	s.t.~~& \text{(\ref{constraint3})-(\ref{constraint6}), (\ref{constraint8})-(\ref{constraint16})},\\
	& \phi^{k,t}_{n},\theta^{k,t}_{m,n},\beta^{k,t}_{m,n},f_{m,n}^t,c_{m,t} \in \{0,1\}.
	\end{array}
\end{equation}

To solve problem (SP2), we need to solve ${\bm{A}, \bm{B}, \bm{C}, \bm{F}}$ given $\delta_m$ first. Since the constraints in (SP2) are independent in each subframe $t$, the problem (SP2) can be decomposed into $T$ concurrent subproblems given $\delta_m$. In subframe $t \in \mathcal{T}_{ON}$, the problem can be given by
\begin{equation}
	\begin{array}{ll}
	\text{SP3}: &\min\limits_{\bm{A}, \bm{B}, \bm{C}, \bm{F}} \left((1-\epsilon) \sum\limits_{m \in \mathcal{M}}\lambda_m (1 - c_{m,t}) - \sum \limits_{m \in \mathcal{M}} \delta_m (c_{m,t} - 1/T)\right),  \\
	s.t.~~& \text{(\ref{constraint3})-(\ref{constraint6}), (\ref{constraint8})-(\ref{constraint12}), (\ref{constraint14}), (\ref{constraint16})},\\
	& \phi^{k,t}_{n},\theta^{k,t}_{m,n},\beta^{k,t}_{m,n},c_{m,t} \in \{0,1\}.
	\end{array}
\end{equation}
In the subframe with the OFF mode, constraints (\ref{constraint12}) and (\ref{constraint14}) will be replaced by constraints (\ref{constraint13}) and (\ref{constraint15}). Since they are in the same function form, the algorithm to solve problem (SP3) can also be used in the subframe with the OFF mode. For brevity, we will only discuss problem (SP3) in the following. 

Then, we will update the Lagrangian multiplier $\delta_m$ according to the following rule until it converges. The update rule can be expressed as
\begin{equation}
\delta_m^{(l+1)} = \left[\delta_m^{(l)} - \eta_{\delta} \left(1- \sum\limits_{t \in \mathcal{T}} c_{m,t}\right) \right]^{+},
\end{equation}
where $[x]^{+} = \max\{x,0\}$, $l$ is the iteration indicator, and $\eta_{\delta}$ is the step size.

\subsubsection{Power Allocation Subproblem}
Given the IoT device association and subchannel allocation $\bm{A},\bm{B}$ and scheduling $\bm{C}$, problem~(P3) can be converted into
\begin{equation}
	\begin{array}{ll}
	\text{SP4}:&\min\limits_{\bm{P}} \sum\limits_{m \in \mathcal{M}} \sum\limits_{t \in \mathcal{T}} \sum\limits_{k \in \mathcal{K} \cup \mathcal{K}^u}\sum\limits_{n \in \mathcal{N} \cup \{0\}} \epsilon p_{m,n}^{k,t} +\sum\limits_{n \in \mathcal{N}} \sum\limits_{t \in \mathcal{T}} \sum\limits_{k \in \mathcal{K}} \epsilon p_{n}^{k,t},  \\
	s.t.~~& \text{(\ref{constraint1}), (\ref{constraint2}),  (\ref{constraint11})-(\ref{constraint16})},\\
	& p_n^{k,t} \geq 0, p_{m,n}^{k,t} \geq 0.
	\end{array}
\end{equation}

Since the power allocation in different subframes is independent, we decouple the problem (SP4) into $T$ subproblems, which can be solved in parallel. For simplicity, we only solve the subprolem in the subframe with the ON mode $t \in \mathcal{T}_{ON}$. The subproblem can be written by
\begin{equation}
	\begin{array}{ll}
	\text{SP5}:&\min\limits_{\bm{P}} \sum\limits_{m \in \mathcal{M}} \sum\limits_{k \in \mathcal{K} \cup \mathcal{K}^u} \sum\limits_{n \in \mathcal{N} \cup \{0\}} \epsilon p_{m,n}^{k,t} +\sum\limits_{n \in \mathcal{N}} \sum\limits_{k \in \mathcal{K}} \epsilon p_{n}^{k,t} \\
	s.t.~~& \text{(\ref{constraint1}), (\ref{constraint11}), (\ref{constraint12}), (\ref{constraint14}), (\ref{constraint16})},\\
	& p_n^{k,t} \geq 0, p_{m,n}^{k,t} \geq 0.
	\end{array}
\end{equation}
The same approach can also be applied to the power allocation in the licensed band. 

\section{Iterative Association, Scheduling, and Power Allocation Algorithm}
\label{Algorithm}
In this section, we will design a low-complexity iterative association, scheduling, and power allocation~(IASPA) algorithm to obtain the suboptimal solutions to problem (P3)\footnote{The IASPA algorithm is performed in a centralized manner. The BS obtains the information of the IoT devices and CUs such as locations and channel conditions, and then performs the IASPA algorithm to obtain the association, scheduling, and power allocation results. After that, the BS will inform the IoT devices and CUs of the results over the control channel.}. Specifically, we solve (SP3) given the power allocation $\bm{P}$, and then solve (SP5) given the IoT device association $\bm{F}$ and scheduling $\bm{A},\bm{B}, \bm{C}$. Finally, we present the overall algorithm. Without loss of generality and for notational brevity, we omit the time index $t$ in the following discussions.

\subsection{IoT Device Association and Scheduling Algorithm}

Note that all the variables in problem (SP3) are integers, and thus, it is an NP-hard problem. To solve this problem effectively, we transform the non-linear constraints into the linear ones. Thus, the problem (P3) is converted to a series of linear integer programming problems, which can be solved by the branch-and-bound algorithm \cite{ED-1966}.

Consider the polynomial constraint (\ref{constraint9}). The following theorem~\cite{DX-2006} shows that the non-linear constraint is equivalent to two linear inequalities.

\begin{theorem}
	Equation (\ref{constraint9}) holds if and only if
	\begin{flalign}
	&- \sum\limits_{n \in \mathcal{Q} \cup \{0\}}(1 - f_{m,n}) + (Q + 1)(1 - c_m) \leq 0,\label{ineq2}\\
	&\sum\limits_{n \in \mathcal{Q} \cup \{0\}}(1 - f_{m,n}) - (1 - c_m) \leq Q.\label{ineq1}
	\end{flalign}
\end{theorem}

\begin{IEEEproof}
If any $f_{m,n} = 1$, then $c_m = 1$. In this case,  (\ref{ineq2}) becomes $c_m \geq  \sum\limits_{n \in \mathcal{Q} \cup \{0\}}\frac{f_{m,n}}{Q+1} >~0$ which implies $c_m = 1$, and (\ref{ineq1}) is redundant. If all $f_{m,n} = 0$, then $c_m = 0$. In such a situation,~(\ref{ineq2}) is redundant, while (\ref{ineq1}) becomes $c_m \leq \sum\limits_{n \in \mathcal{Q} \cup \{0\}}f_{m,n} = 0$ which means $c_m = 0$.
\end{IEEEproof}

Similarly, equation (\ref{constraint10}) holds if and only if
\begin{flalign}
&- \sum\limits_{k \in \mathcal{K}}(1 - \theta^k_{m,n}) + K(1 - f_{m,n}) \leq 0,\\
&\sum\limits_{k \in \mathcal{K}}(1 - \theta^k_{m,n}) - (1 - f_{m,n}) \leq K - 1.
\end{flalign}

In addition, consider that rate constraints (\ref{constraint11}), (\ref{constraint12}), (\ref{constraint14}) and (\ref{constraint16}) are non-convex due to the interference term. To tackle these constraints effectively, we can utilize the a first order Taylor expansion to approximate the data rate into a linear one.

Note that the data rate $R_n^k$ can be rewritten by
\begin{equation}
\begin{array}{ll}
R_n^k \hspace{-3mm}&= \log_2(\sigma^2 + p_{n}^{k}\phi_{n}^{k} |h^{k}_{n}|^2 + \hspace{-3mm}\sum\limits_{m \in \mathcal{S}^M} \hspace{-1mm}\sum\limits_{\begin{subarray}{l}
	n' \in \mathcal{Q},\\
	n' \ne n
	\end{subarray}} {p_{m,{n'}}^{k} \theta_{m,{n'}}^{k} |h^{k}_{m,0}|^2 }) - \log_2(\sigma^2 + \hspace{-3mm} \sum\limits_{m \in \mathcal{S}^M}\hspace{-1mm} \sum\limits_{\begin{subarray}{l}
	n' \in \mathcal{Q},\\
	n' \ne n
	\end{subarray}} {p_{m,{n'}}^{k} \theta_{m,{n'}}^{k} |h^{k}_{m,0}|^2 }),\\
    & \triangleq f_n^k(\phi_n^k,\bm{\theta}_{m,{-n}}^k) - g_n^k(\bm{\theta}_{m,{-n}}^k),
\end{array}
\end{equation}
where $\bm{\theta}_{m,{-n}}^k \triangleq (\bm{\theta}_{m,{n'}}^k)_{m \in \mathcal{M}, n' \in \mathcal{Q} - \{n\} }$ represents all $\theta_{m,{n'}}^k$'s except $\theta_{m,{n}}^k$. Denote $\tilde{\phi}_{n}^k$ and $\tilde{\theta}_{m,n}^k$ by the existing solutions for $\bm{A}$. According to the results in \cite{AGJ-2014}, we have the following approximations:
\begin{flalign}
f_n^k(\phi_n^k,\bm{\theta}_{m,{-n}}^k) &= f_n^k(\phi_n^k,\bm{\tilde{\theta}}_{m,{-n}}^k) + \sum\limits_{m \in \mathcal{S}^M} \hspace{-1mm}\sum\limits_{\begin{subarray}{l}
	n' \in \mathcal{Q},\\
	n' \ne n
	\end{subarray}} a_{m,{n'}}^k (\theta_{m,{n'}}^k - \tilde{\theta}_{m,{n'}}^k),\\
g_n^k(\bm{\theta}_{m,{-n}}^k) &= g_n^k(\bm{\tilde{\theta}}_{m,{-n}}^k) + \sum\limits_{m \in \mathcal{S}^M} \sum\limits_{\begin{subarray}{l}
	n' \in \mathcal{Q},\\
	n' \ne n
	\end{subarray}}b_{m,{n'}}^k (\theta_{m,{n'}}^k - \tilde{\theta}_{m,{n'}}^k)
\end{flalign}
where $a_{m,{n'}}^k$ and $b_{m,{n'}}^k$ are the first order gradient over $\theta_{m,{n'}}^k$ for functions $f_n^k$ and $g_n^k$, respectively. Note that the value of $\theta_{m,{n'}}^k$ can be 0 or 1. Thus, the gradient can be replaced by the difference of function values with $\theta_{m,{n'}}^k = 0$ and $\theta_{m,{n'}}^k = 1$. That is,
\begin{flalign}
a_{m,{n'}}^k &= f_n^k(\tilde{\phi}_n^k,\bm{\tilde{\theta}}_{m,{-\{n,{n'}\}}}^k, \theta_{m,{n'}}^k = 1) - f_n^k(\tilde{\phi}_n^k,\bm{\theta}_{m,{-\{n,{n'}\}}}^k, \theta_{m,{n'}}^k = 0) ,\\
b_{m,{n'}}^k &= g_n^k(\bm{\theta}_{m,{-\{n,{n'}\}}}^k, \theta_{m,{n'}}^k = 1) - g_n^k(\bm{\theta}_{m,{-\{n,{n'}\}}}^k, \theta_{m,{n'}}^k = 0).
\end{flalign}  
Similarly, we can also approximate $f_n^k(\phi_n^k,\bm{\tilde{\theta}}_{m,{-n}}^k)$ by
\begin{equation}
\begin{array}{ll}
f_n^k(\phi_n^k,\bm{\tilde{\theta}}_{m,{-n}}^k) &= \left(f_n^k(1,\bm{\tilde{\theta}}_{m,{-n}}^k) - f_n^k(0,\bm{\tilde{\theta}}_{m,{-n}}^k)\right)\phi_n^k + f_n^k(0,\bm{\tilde{\theta}}_{m,{-n}}^k)\\
&\triangleq a_n^k \phi_n^k + f_n^k(0,\bm{\tilde{\theta}}_{m,{-n}}^k).
\end{array}
\end{equation}
As such, the data rate $R_n^k$ can be approximated by
\begin{equation}
R_n^k = a_n^k \phi_n^k + f_n^k(0,\bm{\tilde{\theta}}_{m,{-n}}^k) - g_n^k(\bm{\tilde{\theta}}_{m,{-n}}^k) + \sum\limits_{m \in \mathcal{S}^M} \hspace{-1mm}\sum\limits_{\begin{subarray}{l}
	n' \in \mathcal{Q},\\
	n' \ne n
	\end{subarray}} (a_{m,{n'}}^k - b_{m,{n'}}^k) (\theta_{m,{n'}}^k - \tilde{\theta}_{m,{n'}}^k)
\end{equation}

Follow the same approach, we can also approximate $R_{m,n}^k$ and $R_{m,n}^{k,u}$ as a linear function of $\theta_{m,n}^k$, $\phi_{n}^k$ and $\beta_{m,n}^k$. Therefore, constraints (\ref{constraint11}), (\ref{constraint12}), (\ref{constraint14}), and (\ref{constraint16}) can be converted into linear constraints, and thus, problem (SP3) is transformed into a linear integer programming, which can be solved by the existing branch-and-bound algorithm.

\subsection{Power Allocation Algorithm}

Due to the existence of the interference term in constraints (\ref{constraint11}), (\ref{constraint12}), (\ref{constraint14}), and (\ref{constraint16}), the problem (SP5) is a non-convex optimization problem. However, we can utilize the SCA technique~\cite{JJ-2009} to approximate the non-convex functions into a series of convex ones. As such, the non-convex problem can be transformed into a convex one and solved by the Lagrange dual technique~\cite{SL-2004}.

\subsubsection{Successive Convex Approximation}
According to the results in \cite{JJ-2009}, the data rate $R_{n}^{k}$ can be approximated by
\begin{equation}
\tilde{R}_{n}^{k} = d_{n}^{k} \log(\gamma_{n}^{k}) + e_{n}^{k},
\end{equation}
where
\begin{equation}
d_{n}^{k} = \frac{\tilde{\gamma}_{n}^{k}}{1 + \tilde{\gamma}_{n}^{k}}, ~~e_{n}^{k} = \frac{1}{\log 2}\log(1 + \tilde{\gamma}_{n}^{k}) - d_{n}^{k}\log(\tilde{\gamma}_{n}^{k}).
\end{equation}
Here, $\tilde{\gamma}_{n}^{k}$ is calculated based on the current power allocation result. Likewise, $R_{m,n}^k$ and $R_{m,n}^{k,u}$ can be also expressed by
\begin{equation}
\tilde{R}_{m,n}^{k} = d_{m,n}^{k} \log(\gamma_{m,n}^{k}) + e_{m,n}^{k},~~\tilde{R}_{m,n}^{k,u} = d_{m,n}^{k,u} \log(\gamma_{m,n}^{k,u}) + e_{m,n}^{k,u}.
\end{equation}

However, $\log(\tilde{\gamma}_{n}^{k})$, $\log(\tilde{\gamma}_{m,n}^{k})$, and $\log(\tilde{\gamma}_{m,n}^{k,u})$ are still non-convex. Therefore, we can introduce the transformations of $p_{m,n}^k = \exp(\hat{p}_{m,n}^k)$ and $p_{n}^k = \exp(\hat{p}_{n}^k)$. As such, problem (SP5) can be converted into the following problem:
\begin{subequations}
	\begin{align}
	\text{SP6}:\min\limits_{\bm{\hat{P}}} & \sum\limits_{m \in \mathcal{M}}  \sum\limits_{k \in \mathcal{K} \cup \mathcal{K}^u} \sum\limits_{n \in \mathcal{Q} \cup \{0\}} \epsilon \exp(\hat{p}_{m,n}^k) +\sum\limits_{n \in \mathcal{N}}  \sum\limits_{k \in \mathcal{K}} \epsilon \exp(\hat{p}_{n}^k) \\
	s.t.~~& \sum\limits_{k \in \mathcal{K}} \exp(\hat{p}_{m,0}^k) + \sum\limits_{k \in \mathcal{K} \cup \mathcal{K}^u} \sum \limits_{n \in \mathcal{Q}} \exp(\hat{p}_{m,n}^k) \leq P^{iot}, \sum\limits_{k \in \mathcal{K} }\exp(\hat{p}_{n}^k) \leq P^c, \forall m \in \mathcal{M}, n \in \mathcal{N},\label{co1}\\
	& \sum\limits_{k \in \mathcal{K}} (d_{m,0}^{k} \log(\gamma_{m,0}^k) + e_{m,0}^k) \geq d_m c_{m,t}, \forall m \in \mathcal{S}^C,\label{co2}\\
	&\sum\limits_{n \in \mathcal{Q}}\left(\sum\limits_{k \in \mathcal{K}}  (d_{m,n}^{k} \log(\gamma_{m,n}^k) + e_{m,n}^k) + \sum\limits_{k \in \mathcal{K}^u}  (d_{m,n}^{k,u} \log(\gamma_{m,n}^{k,u}) + e_{m,n}^{k,u})\right) \geq d_m c_{m,t}, \forall m \in \mathcal{S}^M,\label{co3}\\
	&\sum\limits_{k \in \mathcal{K}} (d_{n}^{k} \log(\gamma_{n}^k) + e_{n}^k) \geq \sum\limits_{m \in \mathcal{S}^M} \sum\limits_{k \in \mathcal{K}}(d_{m,n}^{k} \log(\gamma_{m,n}^k) + e_{m,n}^k) \nonumber\\
	&~~~~~~~~~~~~~~~~~~~~~~~~~+ \sum\limits_{m \in \mathcal{S}^M} \sum\limits_{k \in \mathcal{K}^u}(d_{m,n}^{k,u} \log(\gamma_{m,n}^{k,u}) + e_{m,n}^{k,u}), \forall n \in \mathcal{Q},\label{co4}\\
	&\sum\limits_{k \in \mathcal{K}}(d_{n}^{k} \log(\gamma_{n}^k) + e_{n}^k) \geq R_{min}, \forall n \in \mathcal{N} \backslash \mathcal{Q},\label{co5}
	\end{align}
\end{subequations}
where
\begin{flalign}
	& \log(\gamma_{n}^{k}) = \hat{p}_{n}^k + \log(|h_{n}^k|^2\phi_{n}^k) - \log(\sigma^2 + \sum \limits_{m \in \mathcal{S}^M} \sum \limits_{\begin{subarray}{l}
		n' \in \mathcal{Q},\\
		n' \ne n
		\end{subarray}} {\exp(\hat{p}_{m,{n'}}^{k}) \theta_{m,{n'}}^{k,t} |h^{k}_{m,0}|^2 }),\\	
	&\log(\gamma_{m,0}^{k}) = \hat{p}_{m,0}^k + \log(|h_{m,0}^k|^2\phi_{m,0}^k) - \log(\sigma^2 + \sum \limits_{m' \in \mathcal{S}^M} \sum \limits_{n \in \mathcal{Q}} {\exp(\hat{p}_{{m'},n}^{k}) \theta_{{m'},n}^{k} |h^{k}_{{m'},0}|^2 }),\\
	&\log(\gamma_{m,n}^{k}) = \hat{p}_{m,n}^k + \log(|h_{m,n}^k|^2\phi_{m,n}^k)\nonumber\\
	&~~~~~~~~~~~ - \log(\sigma^2 + \sum \limits_{\begin{subarray}{l}
		n' \in \mathcal{N},\\
		n' \ne n
		\end{subarray}} \exp(\hat{p}_{n'}^k)\phi_{n'}^k|h_{{n'},n}^k|^2 + \sum\limits_{\begin{subarray}{l}
		m' \in \mathcal{M},\\
		m' \ne m
		\end{subarray}}\sum \limits_{\begin{subarray}{l}
		n' \in \mathcal{Q} \cup \{0\}\\
		n' \ne n
		\end{subarray}} {\exp(\hat{p}_{{m'},{n'}}^{k}) \theta_{{m'},{n'}}^{k} |h^{k}_{{m'},n}|^2 }),\\
	&\log(\gamma_{m,n}^{k,u}) = \hat{p}_{m,n}^{k} + \log(|h_{m,n}^{k,u}|^2\beta_{m,n}^{k,u}) - \log(\sigma^2 + \sum\limits_{\begin{subarray}{l}
		m' \in \mathcal{M},\\
		m' \ne m
		\end{subarray}}\sum \limits_{\begin{subarray}{l}
		n' \in \mathcal{Q},\\
		n' \ne n
		\end{subarray}} {\exp(\hat{p}_{{m'},{n'}}^{k}) \beta_{{m'},{n'}}^{k} |h^{k,u}_{{m'},n}|^2 + I_{m,n}^{k,u}}).
\end{flalign}
Since the log-sum-exp function is convex \cite{SL-2004}, problem (SP6) is a standard convex minimization problem, which can be solved by the Lagrange dual technique. Once the optimal solution is obtained, we may transform back with $p_{m}^k = \exp(\hat{p}_{m}^k)$ and $p_{n}^k = \exp(\hat{p}_{n}^k)$. To find the optimal power allocation, we need to iteratively update the approximation parameters according to the allocated power.

\subsubsection{Lagrange Dual Technique}

Define $\bm{\mu}$, $\bm{\nu}$, $\bm{\kappa}$, $\bm{\xi}$ and $\bm{\chi}$ are Lagrange multiplier vectors corresponding to constraints (\ref{co1}), (\ref{co2}), (\ref{co3}), (\ref{co4}) and (\ref{co5}), respectively. Therefore, the Lagrangian can be written as
\begin{equation}
\begin{array}{ll}
&L(\bm{\hat{P}}, \bm{\mu}, \bm{\nu}, \bm{\kappa}, \bm{\xi},\bm{\chi})  = \sum\limits_{m \in \mathcal{M}}  \sum\limits_{k \in \mathcal{K} \cup \mathcal{K}^u} \sum\limits_{n \in \mathcal{Q} \cup \{0\}} \epsilon \exp(\hat{p}_{m,n}^k) +\sum\limits_{n \in \mathcal{N}}  \sum\limits_{k \in \mathcal{K}} \epsilon \exp(\hat{p}_{n}^k)\\
&~~~+ \sum \limits_{m \in \mathcal{M}} \mu_m \left(P^{iot} - \sum\limits_{k \in \mathcal{K}} \exp(\hat{p}_{m,0}^k) - \sum\limits_{k \in \mathcal{K} \cup \mathcal{K}^u} \sum \limits_{n \in \mathcal{Q}} \exp(\hat{p}_{m,n}^k) \right)  + \sum \limits_{n \in \mathcal{N}} \mu_n \left(P^{c} - \sum\limits_{k \in \mathcal{K}} \exp(\hat{p}_{n}^k) \right)\\
&~~~+ \sum \limits_{m \in \mathcal{S}^C} \nu_m\left(\sum \limits_{k \in \mathcal{K}}(d_{m,0}^{k} \log(\gamma_{m,0}^k) + e_{m,0}^k) - d_m c_{m,t}\right)\\
&~~~+ \sum \limits_{m \in \mathcal{S}^M} \kappa_m \left(\sum\limits_{n \in \mathcal{Q}}(\sum\limits_{k \in \mathcal{K}}  (d_{m,n}^{k} \log(\gamma_{m,n}^k) + e_{m,n}^k) + \sum\limits_{k \in \mathcal{K}^u}  (d_{m,n}^{k,u} \log(\gamma_{m,n}^{k,u}) + e_{m,n}^{k,u})) - d_m c_{m,t} \right)\\
&~~~+\sum \limits_{n \in \mathcal{Q}} \xi_n \left(\sum\limits_{k \in \mathcal{K}} (d_{n}^{k} \log(\gamma_{n}^k) + e_{n}^k) - \sum\limits_{m \in \mathcal{S}^M} \sum\limits_{k \in \mathcal{K}}(d_{m,n}^{k} \log(\gamma_{m,n}^k) + e_{m,n}^k)\right.\\
&~~~\left.-\sum\limits_{m \in \mathcal{S}^M} \sum\limits_{k \in \mathcal{K}^u}(d_{m,n}^{k,u} \log(\gamma_{m,n}^{k,u}) + e_{m,n}^{k,u})\right) + \sum\limits_{n \in \mathcal{N} \backslash \mathcal{Q}}\chi_n\left(\sum\limits_{k \in \mathcal{K}}(d_{n}^{k} \log(\gamma_{n}^k) + e_{n}^k) - R_{min}\right).
\end{array}
\end{equation}
and the dual problem can be given by
\begin{equation}
\max \limits_{\bm{\mu},\bm{\nu},\bm{\kappa},\bm{\xi},\bm{\chi} \succeq \bm{0}} \min \limits_{\bm{\hat{P}}} L(\bm{\hat{P}}, \bm{\mu}, \bm{\nu}, \bm{\kappa}, \bm{\xi},\bm{\chi}).
\end{equation}

We can solve the dual problem iteratively by decomposing it into two subproblems. The master subproblem is the maximization of $\bm{\mu}$, $\bm{\nu}$, $\bm{\kappa}$, $\bm{\xi}$, and $\bm{\chi}$. The slave subproblem is to maximize $\bm{\hat{P}}$ given $\bm{\mu}$, $\bm{\nu}$, $\bm{\kappa}$, $\bm{\xi}$, and $\bm{\chi}$. In the following, we will discuss them in detail.

\textbf{Slave Subproblem:} According to the Karush-Kuhn-Tucker~(KKT) conditions, we can obtain the optimal power allocation by equaling the first derivative of the Lagrangian function $L(\bm{\hat{P}}, \bm{\mu}, \bm{\nu}, \bm{\kappa}, \bm{\xi}, \bm{\chi})$ over $\bm{\hat{P}}$ to 0. Therefore, we have
\begin{equation}
\begin{array}{ll}
\exp(\hat{p}_n^k) &= \left.{d_{n}^k\xi_n}\middle/{\left(\mu_n - \epsilon +\sum \limits_{m \in \mathcal{S}^M}\sum \limits_{\begin{subarray}{l}
		n' \in \mathcal{Q},\\
		n' \ne n
		\end{subarray}}(\kappa_m d_{m,{n'}}^k - \xi_{n'}d_{m,{n'}}^k)\frac{\gamma_{m,{n'}}^k\phi_n^k\theta_{m,{n'}}^k|h_{n,{n'}}^k|^2}{\exp(\hat{p}_{m,{n'}}^k)|h_{m,{n'}}^k|^2}\right)}\right.,\\
&~~~~~~~~~~~~~~~~~~~~~~~~~~~~~~~~~~~~~~~~~~~~~~~~~~~~~~~~~~~~~~~~~~~~~~\forall n \in \mathcal{Q}, k \in \mathcal{K},
\end{array}
\end{equation}
\begin{equation}
\exp(\hat{p}_n^k) = \left.{d_{n}^k\chi_n}\middle/{\left(\mu_n - \epsilon +\sum \limits_{m \in \mathcal{S}^M}\sum \limits_{\begin{subarray}{l}
		n' \in \mathcal{Q},\\
		\end{subarray}}\kappa_m d_{m,{n'}}^k \frac{\gamma_{m,{n'}}^k\phi_n^k\theta_{m,{n'}}^k|h_{n,{n'}}^k|^2}{\exp(\hat{p}_{m,{n'}}^k)|h_{m,{n'}}^k|^2}\right)}\right.,\forall n \in \mathcal{N} \backslash \mathcal{Q}, k \in \mathcal{K},
\end{equation}
\begin{equation}
\begin{array}{ll}
&\hspace{-2mm}\exp(\hat{p}_{m,0}^k) = \left.{d_{m,0}^k\nu_m}\middle/{\left(\mu_m \hspace{-1mm}-\hspace{-1mm} \epsilon \hspace{-1mm}+\hspace{-1mm}\sum \limits_{\begin{subarray}{l}
		m' \in \mathcal{S}^M,\\
		m' \ne m
		\end{subarray}}\sum \limits_{{n} \in \mathcal{Q}}(\kappa_{m'} d_{{m'},{n}}^k \hspace{-1mm}-\hspace{-1mm} \xi_{n}d_{{m'},{n}}^k)\frac{\gamma_{{m'},{n}}^k\theta_{m,0}^k|h_{m,{n}}^k|^2\theta_{{m'},{n}}^k}{\exp(\hat{p}_{{m'},{n}}^k)|h_{{m'},{n}}^k|^2}\right)}\right., \\
&~~~~~~~~~~~~~~~~~~~~~~~~~~~~~~~~~~~~~~~~~~~~~~~~~~~~~~~~~~~~~~~~~~~~~~~~~~~~~~\forall m \in \mathcal{S}^C, k \in \mathcal{K},
\end{array}
\end{equation}
\begin{equation}
\begin{array}{ll}
&\exp(\hat{p}_{m,n}^k) = \left.{\left(\kappa_m d_{m,n}^k - \xi_n d_{m,n}^k\right)}\middle/\left(
	\mu_m - \epsilon +\sum \limits_{\begin{subarray}{l}
		m' \in \mathcal{S}^C,\\
		m' \ne m
		\end{subarray}}\nu_{m'} d_{{m'},0}^k \frac{\gamma_{{m'},{0}}^k\theta_{m,n}^k\theta_{{m'},{0}}^k|h_{m,{0}}^k|^2}{\exp(\hat{p}_{{m'},{0}}^k)|h_{{m'},{0}}^k|^2} + \right.\right.\\
	&\left.\left.\sum \limits_{\begin{subarray}{l}
		n' \in \mathcal{Q},\\
		n' \ne n
		\end{subarray}}\xi_{n'} d_{{n'}}^k \frac{\gamma_{{n'}}^k\theta_{m,n}^k\phi_{{n'}}^k|h_{m,{0}}^k|^2}{\exp(\hat{p}_{{n'}}^k)|h_{{n'}}^k|^2} + \sum \limits_{\begin{subarray}{l}
		n' \in \mathcal{N} \backslash \mathcal{Q},\\
		n' \ne n
		\end{subarray}}\chi_{n'} d_{{n'}}^k \frac{\gamma_{{n'}}^k\theta_{m,n}^k\phi_{{n'}}^k|h_{m,{0}}^k|^2}{\exp(\hat{p}_{{n'}}^k)|h_{{n'}}^k|^2} + \right.\right.\\
	&\left.\left. \sum \limits_{\begin{subarray}{l}
		m' \in \mathcal{S}^M,\\
		m' \ne m
		\end{subarray}}\sum \limits_{\begin{subarray}{l}
		n' \in \mathcal{Q},\\
		n' \ne n
		\end{subarray}} (\kappa_{m'}d_{{m'},{n'}}^{k} \hspace{-1mm}-\hspace{-1mm} \xi_{n'}d_{{m'},{n'}}^{k})\frac{\gamma_{{m'},{n'}}^{k}\theta_{m,n}^k\theta_{{m'},{n'}}^k|h_{m,{n'}}^k|^2}{\exp(\hat{p}_{{m'},{n'}}^k)|h_{{m'},{n'}}^{k}|^2}
	\right)\right.,\forall m \in \mathcal{S}^M, n \in \mathcal{Q}, k \in \mathcal{K},
\end{array}
\end{equation}
\begin{equation}
\begin{array}{ll}
\exp(\hat{p}_{m,n}^k) &= \left.{\left(\kappa_m d_{m,n}^{k,u} - \xi_n d_{m,n}^{k,u}\right)}\middle/\left(\mu_m - \epsilon + \sum \limits_{\begin{subarray}{l}
			m' \in \mathcal{S}^M,\\
			m' \ne m
			\end{subarray}}\sum \limits_{\begin{subarray}{l}
			n' \in \mathcal{N},\\
			n' \ne n
			\end{subarray}} (\kappa_{m'}d_{{m'},{n'}}^{k,u} - \xi_{n'}d_{{m'},{n'}}^{k,u})\right.\right.\\
		&~~~~~~~~~~~~~~~~~~~~\left.\left.\frac{\gamma_{{m'},{n'}}^{k,u}\beta_{m,n}^k\beta_{{m'},{n'}}^k|h_{m,{n'}}^{k,u}|^2}{\exp(\hat{p}_{{m'},{n'}}^k)|h{{m'},{n'}}^{k,u}|^2}\right)\right., \forall m \in \mathcal{S}^M, n \in \mathcal{N}, k \in \mathcal{K}^u.
\end{array}
\end{equation}

As such, we can obtain the optimal power allocation results for all the scheduled IoT devices.

\textbf{Master Subproblem:} Once the optimal power allocation is achieved, the solution of the dual problem can be obtained by a subgradient method \cite{S-2013} as follows:

\begin{flalign}
\mu_{n}^{(l + 1)} &= \left[\mu_{n}^{(l)} - \eta_{\mu}\left(P^{c} - \sum\limits_{k \in \mathcal{K}} \exp(\hat{p}_{n}^k) \right) \right]^{+}, \forall n \in \mathcal{Q},\\
\mu_{m}^{(l + 1)} &= \left[\mu_{m}^{(l)} - \eta_{\mu}\left(P^{iot} - \sum\limits_{k \in \mathcal{K}} \exp(\hat{p}_{m,0}^k) - \sum\limits_{k \in \mathcal{K} \cup \mathcal{K}^u} \sum \limits_{n \in \mathcal{Q}} \exp(\hat{p}_{m,n}^k) \right) \right]^{+},\forall m \in \mathcal{M},\\
\nu_m^{(l + 1)} &= \left[\nu_{m}^{(l)} - \eta_{\nu} \left(\sum \limits_{k \in \mathcal{K}}(d_{m,0}^{k} \log(\gamma_{m,0}^k) + e_{m,0}^k) - d_m c_{m,t}\right) \right]^{+},\forall m \in \mathcal{S}^C,\\
\kappa_m^{(l + 1)} &= \left[\kappa_m^{(l)} - \eta_{\kappa} \left(\sum\limits_{n \in \mathcal{Q}}(\sum\limits_{k \in \mathcal{K}}  (d_{m,n}^{k} \log(\gamma_{m,n}^k) + e_{m,n}^k)  \right.\right.\nonumber\\ 
 &\phantom{=\;\;}\left.\left. +\sum\limits_{k \in \mathcal{K}^u}  (d_{m,n}^{k,u} \log(\gamma_{m,n}^{k,u}) + e_{m,n}^{k,u})) - d_m c_{m,t} \right) \right]^{+}, \forall m \in \mathcal{S}^M,\\
 \xi_n^{(l + 1)} & = \left[\xi_n^{(l)} - \eta_{\xi} \left(\sum\limits_{k \in \mathcal{K}} (d_{n}^{k} \log(\gamma_{n}^k) + e_{n}^k) - \sum\limits_{m \in \mathcal{M}} \sum\limits_{k \in \mathcal{K}}(d_{m,n}^{k} \log(\gamma_{m,n}^k) + e_{m,n}^k)\right.\right.\nonumber\\
 &\phantom{=\;\;}\left.\left.-\sum\limits_{m \in \mathcal{M}} \sum\limits_{k \in \mathcal{K}^u}(d_{m,n}^{k,u} \log(\gamma_{m,n}^{k,u}) + e_{m,n}^{k,u})\right)\right]^{+}, \forall n \in \mathcal{Q},\\
 \chi_n^{(l + 1)} &= \left[\chi_n^{(l)} - \eta_{\chi}\left(\sum\limits_{k \in \mathcal{K}}(d_{n}^{k} \log(\gamma_{n}^k) + e_{n}^k) - R_{min}\right)\right]^{+}, \forall n \in \mathcal{N} \backslash \mathcal{Q},
\end{flalign}
where $l$ is the iteration indicator, and $\eta_{\mu}$, $\eta_{\nu}$, $\eta_{\kappa}$, $\eta_{\xi}$ and $\eta_{\chi}$ are sufficiently small step sizes to guarantee convergence. As such, the optimal solution for the dual problem when the multiplier vectors converge.

\subsection{Overall Algorithm}
Based on the results presented in the previous two subsections, we propose an overall iterative algorithm, i.e., IASPA algorithm for problem (P3) by applying the alternating optimization method. Initially, we set the power for each link are allocated to all the subchannels equally. Then the IoT device association and scheduling variables $\bm{A}$, $\bm{B}$, $\bm{C}$, and $\bm{F}$ are optimized by solving subproblem (SP3) while keeping the allocated power fixed. After obtaining the association and scheduling results, we will optimize the power allocation $\bm{P}$ on these scheduled IoT devices by solving subproblem (SP5). Those scheduled IoT will utilize the optimized transmit power as the initial power in the next iterations. Define $O(\bm{A},\bm{B},\bm{C},\bm{F},\bm{P})$ as the value of the objective function in problem (P3). The two subproblems will be solved alternatively until the value difference of the objective functions between two successive iterations is less than a predefined threshold $\pi$, i.e., $|O(\bm{A}^{l + 1},\bm{B}^{l + 1},\bm{C}^{l + 1},\bm{F}^{(l + 1)},\bm{P}^{l + 1}) - O(\bm{A}^{l},\bm{B}^{l},\bm{C}^{l},\bm{F}^{(l)},\bm{P}^{l})| \leq \pi$, where $l$ denotes the iteration. The details of IASPA algorithm are summarized in Algorithm \ref{algorithm}. 

\begin{algorithm}[!t]
	\caption{The Procedure of IASPA Algorithm}\label{algorithm}
	\begin{algorithmic}[1] 
		\Require
		The set of IoT devices $\mathcal{M}$; 
		The set of CUs $\mathcal{N}$; 
		The set of idle CUs $\mathcal{Q}$;
		The number of licensed subchannels $K$ and unlicensed subchannels $K^u$; 
		The number of subframes for one cycle $T$; 
		\Ensure
		IoT association and scheduling matrices $\bm{A}$, $\bm{B}$, $\bm{C}$, and $\bm{F}$;
		Transmit power matrix $\bm{P}$;
		\State Initialize transmit power $\bm{P}_{int}^{0}$ by setting $p_{m,n}^k = P^{iot}/(K + K^u)$ and $p_n^k = P^{c}/K$, and $l = 0$;
		\Repeat
		\State $l = l + 1$;
		\State Solve subproblem (SP3) given the initial power allocation $\bm{P}_{int}^{l - 1}$, and obtain the association and scheduling result $\bm{A}^{l}$, $\bm{B}^{l}$,  $\bm{C}^{l}$, and $\bm{F}^{l}$;
		\State Solve subproblem (SP5) given the association and scheduling result $\bm{A}^{l}$, $\bm{B}^{l}$, $\bm{C}^{l}$, and $\bm{F}^{l}$, and then obtain the power allocation result $\bm{P}^{l}$;
		\State Use the solved power allocation result to replace the transmit power of the scheduled IoT devices in $\bm{P}_{int}^{l}$ and obtain an updated initial transmit power matrix $\bm{P}_{int}^{1 + 1}$ for the next iteration;
		\Until{$|O(\bm{A}^{l},\bm{B}^{l},\bm{C}^{l},\bm{F}^{l},\bm{P}^{l}) - O(\bm{A}^{l - 1},\bm{B}^{l-1},\bm{C}^{l-1},\bm{F}^{l-1}\bm{P}^{l-1})| \leq \pi$;}
	\end{algorithmic}
\end{algorithm}

\section{Performance Analysis}\label{Discussion}

In this section, we analyze the effectiveness and efficiency of the proposed algorithm, and remark some key properties of the IoT-U network. In the first part, the convergence and the complexity of the proposed IASPA algorithm is proved. Then, we discuss the IoT devices association and the offloading gain of unlicensed spectrum.

\subsection{Convergence}
First, in the IoT association and scheduling subproblem, we can obtain a better association and scheduling solution given $\bm{P}^{l}$. Therefore, we have 
\begin{equation}\label{ine1}
O(\bm{A}^{l + 1},\bm{B}^{l+1},\bm{C}^{l+1},\bm{F}^{l+1},\bm{P}^{l}) \leq O(\bm{A}^{l},\bm{B}^{l},\bm{C}^{l},\bm{F}^{l},\bm{P}^{l}).
\end{equation}

Second, given the association and scheduling $\bm{A}^{l + 1}$, $\bm{B}^{l+1}$, $\bm{C}^{l+1}$,$\bm{F}^{l+1}$, we minimize the allocated power, and thus, the following inequality holds:
\begin{equation}\label{ine2}
O(\bm{A}^{l + 1},\bm{B}^{l+1},\bm{C}^{l+1},\bm{F}^{l+1},\bm{P}^{l + 1}) \leq O(\bm{A}^{l + 1},\bm{B}^{l + 1},\bm{C}^{l + 1},\bm{F}^{l+1}\bm{P}^{l}).
\end{equation} 

Based the inequalities (\ref{ine1}) and (\ref{ine2}), we can obtain
\begin{equation}
O(\bm{A}^{l + 1},\bm{B}^{l+1},\bm{C}^{l+1},\bm{F}^{l+1},\bm{P}^{l + 1}) \leq O(\bm{A}^{l},\bm{B}^{l},\bm{C}^{l},\bm{F}^{l},\bm{P}^{l}),
\end{equation} 
which implies that the objective value of problem (P3) is non-decreasing after each iteration of the ISAPA Algorithm. Since the objective value of problem (P3) is upper bounded by a finite value, the proposed ISAPA algorithm is guaranteed to converge.

\subsection{Complexity}

In the IoT device association and scheduling subproblem, we approximate it as a sequence of integer linear programming problems. Based on the results in \cite{A-1998}, the integer linear programming problem is a polynomial computational complexity related to the number of variables, i.e., $M(N+1)(K+1)T + MNK^uT$. 

In power allocation subproblem, we approximate it as a series of convex problems and solve it by the primary-dual interior point method. According to the results in \cite{SL-2004}, the scale of the computational complexity of each convex problem is $O(\sqrt{Z}\log(\frac{1}{\tau}))$, where $Z = (M + 1)(N + 1)$ is the number of constraints and $\tau$ is the tolerance threshold for convergence.

\subsection{Association}\label{Association}

\begin{proposition}\label{proposition}
	Define the data rate of a CU is $R$. The number of the IoT devices associating with this CU cannot exceed $\min\{K - 1, R/d\}$. 
\end{proposition}

\begin{IEEEproof}
	On the one hand, note that the associated IoT devices cannot share the same spectrum with the CU, and thus, these IoT devices can only utilize at most $K - 1$ licensed subchannels. Besides, due to the CA requirement, each IoT device need to be allocated to at least one licensed subchannel. Therefore, the number of the associated IoT devices cannot be larger than $K - 1$. On the other hand, each IoT device has $d$ data traffic to transmit, and therefore, the CU can support at most $R/d$ IoT devices.
\end{IEEEproof}

\begin{remark}\label{remark}
	The IoT devices with high data traffic will tend to associate with the BS while those with low data traffic will tend to associate with the CUs.
\end{remark}

\begin{IEEEproof}
When the data traffic is high, for example, the data rate for the cellular link can support the aggregated data from only one IoT device, the data rate in the cellular link will be the bottleneck. Therefore, the IoT device will associate with the BS directly since it will not bring the interference. However, when the data traffic is low, the cellular link can support the data aggregated from more than one IoT devices. Even the IoT device is located close to the BS, the IoT device will still associate with a neighboring CU because a subchannel can support more than one IoT devices in this case. 	
\end{IEEEproof}

\subsection{Offloading Gain}

\begin{proposition}\label{schedule}
	The number of scheduled IoT devices in the IoT-U network is no less than that using the licensed scheme where only the licensed spectrum is utilized.
\end{proposition}
\begin{IEEEproof}
	Note that the number of scheduled IoT devices using the licensed scheme is equivalent to solve (P1) by setting $\beta_{m,n}^{k,t} = 0$. Therefore, the solution obtained by the licensed scheme is also a feasible solution to (P1). Since the proposed unlicensed scheme is to find the optimal solution of (P1), the value of the objective obtained by the proposed unlicensed scheme is no less than that obtained by the licensed scheme. 
\end{IEEEproof}

\begin{remark}
	Define $R_L$ as the expectation of the data rate for a cellular link over one subchannel using the licensed scheme and $R_U$ as that using the scheme where there does not exist a D2D link over the licensed spectrum. The number of scheduled IoT devices in the IoT-U network can increase $R_U/R_L - 1$ at most compared to the licensed scheme.
\end{remark}
\begin{IEEEproof}
    In the unlicensed scheme, the IoT devices associated with a CU will transmit with the most of power over the unlicensed band. Therefore, the less transmit power over the licensed band alleviates the interference to the cellular links, which brings the offloading gain. For this reason, the IoT-U network can schedule at most $R_U/d$ where $d$ is the expectation of the data traffic for an IoT device when the interference is neglected, while the licensed scheme schedules $R_L/d$ IoT devices. Thus, the IoT-U network can schedule at most $R_U/R_L - 1$ more IoT devices than the licensed scheme.
\end{IEEEproof}

\section{Simulation Results}
\label{sec:simulation}

\begin{table}
	\centering
	\caption{Parameters for Simulation} \label{parameters}
	\begin{tabular}{|p{6cm}|p{3cm}|}
		\hline \textbf{Parameters} & \textbf{Values}\\
		\hline
		\hline Number of WUs & 3\\
		\hline WU's transmit power $P^w$ & 23 dBm\\
		\hline Transmission bandwidth & 180 kHz \\
		\hline Interval for a time slot & 100 ms \\
		\hline Number of time slots for a cycle $T$ & 6\\
		\hline Duty cycle percentage $T_{ON}/T$ & 0.5 \\
		\hline Carrier frequency of licensed band & 1.9 GHz\\
		\hline Carrier frequency of unlicensed band & 5 GHz\\
		\hline Noise figure & 5 dB \\
		\hline Error tolerance level $\pi$ & 0.1 \\
		\hline
	\end{tabular}
\end{table}

In this section, we evaluate the performance of our proposed IoT-U scheme. For comparison,
the following schemes are also performed:
\begin{itemize}
	\item Traditional cellular (TC) networks: All IoT devices transmit their sensory data to the BS in cellular mode.
	\item Cluster based IoT networks over licensed spectrum (IoT-L): The IoT devices adopt the cluster framework where only the licensed spectrum can be utilized.
	\item Decoupled association, scheduling, and power allocation (DASPA) scheme: The IoT devices association and scheduling are performed first, and then the transmission power on each subchannel is allocated to minimize the total energy consumption.
	\item Matching based association, scheduling, and power allocation (MASPA) scheme \cite{TNWLDTC-2017}: The IoT devices association is formulated as one-to-one matching, the scheduling is many-to-one matching, and the power allocation is determined by geometric programming.
\end{itemize}

In our simulation, we set the radius of the cell as 500 m. The number of CUs, idle CUs, and IoT devices are set as $N = 15$, $Q = 10$, and $M = 300$, respectively. These CUs and IoT devices are uniformly distributed in the cell. The maximum transmit powers of each CU  and each IoT device are set as $P^c = 23$ dBm and $P^{iot} = 13$ dBm, respectively. For simplicity, we assume that the data requirement for different IoT devices are the same, i.e., $d_1 = \ldots = d_M = d$. To satisfy the requirements for different applications, the value of $d$ varies from 200 kbps to 1 Mbps. The minimum data rate requirement for a CU is set as $R_{min} = 1$ Mbps. We set the numbers of licensed and unlicensed subchannels as $K = 9\sim15$ and $K^u = K$. The small-scale fading is modeled as the Rayleigh fading. We adopt the path loss model in~\cite{3GPP-2013} where the decay factor $\alpha = 3.76$ and power gains factor $G = -17.7$ dB. Other simulation parameters are set according to \cite{HYL-2017}, as listed in Table \ref{parameters}.

\begin{figure}
\centering
\subfigure[]{
\label{Figure1}
\includegraphics[height=2.2in]{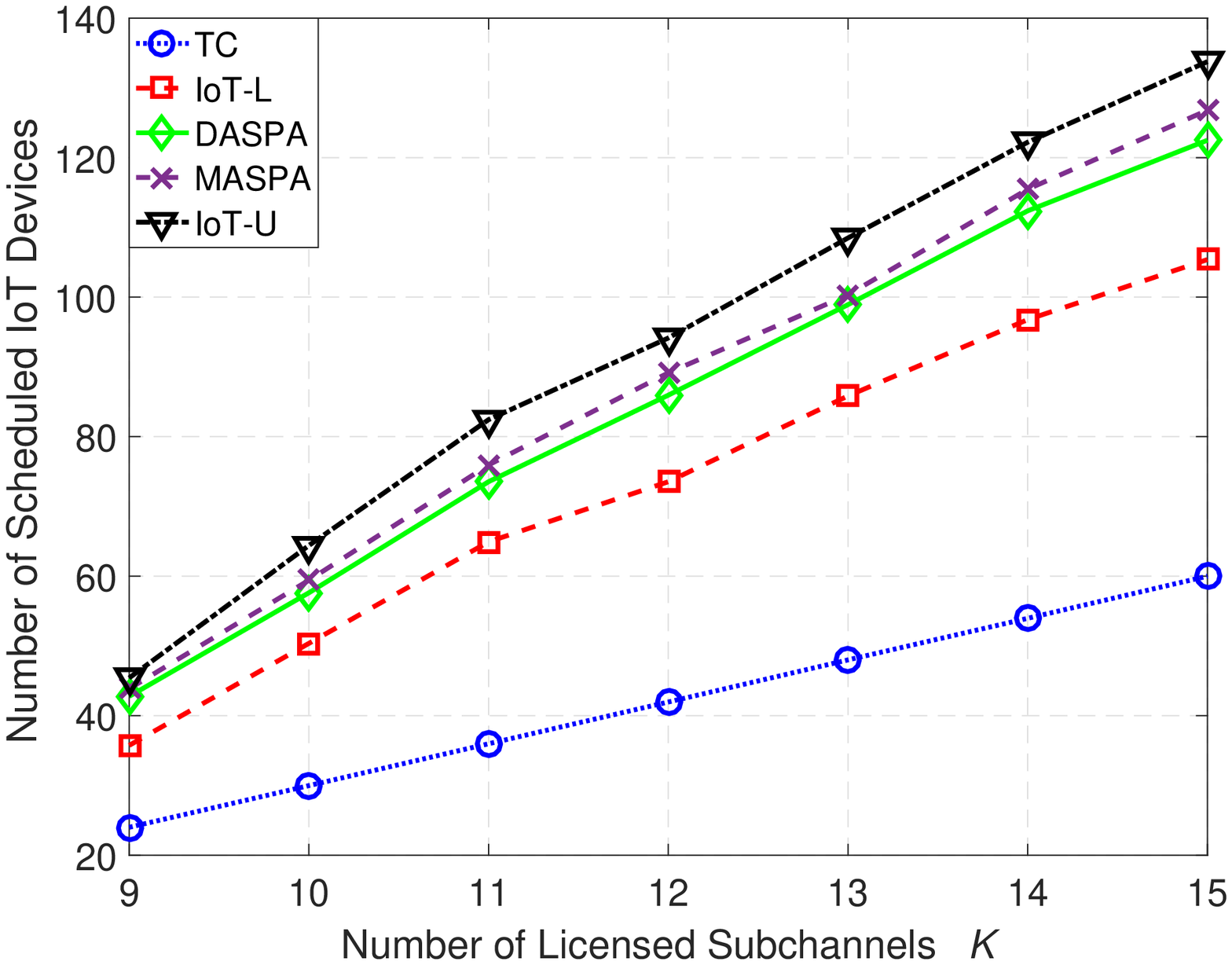}}
\subfigure[]{
\label{Figure2}
\includegraphics[height=2.2in]{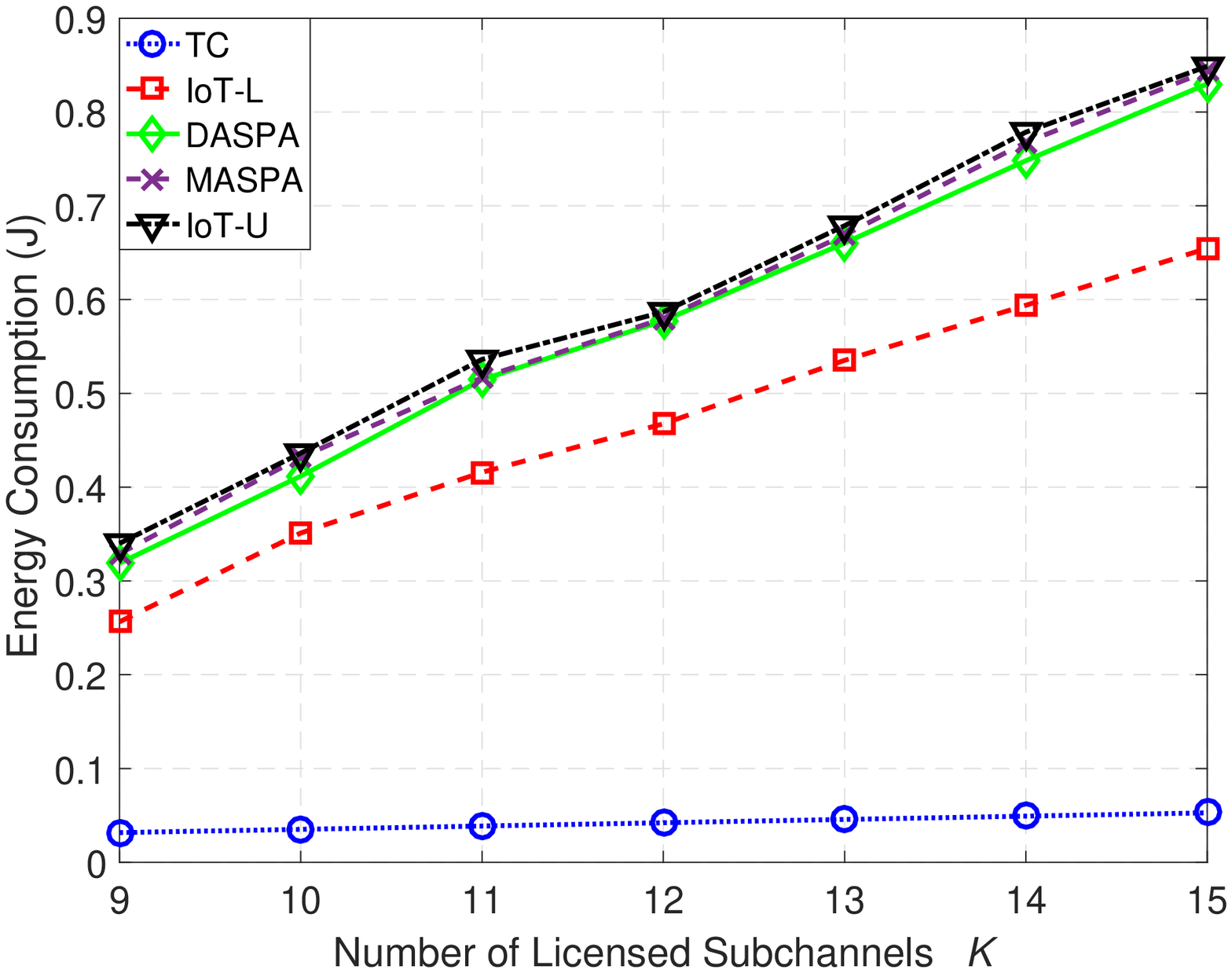}}
\caption{(a) Number of scheduled IoT devices v.s. number of licensed subchannels; (b) Energy Consumption v.s. number of licensed subchannels with $d = 500$ kbps.}\label{Figure12}
\vspace{-5mm}
\end{figure}

Fig.~\ref{Figure1} illustrates the number of scheduled IoT devices v.s. the number of licensed subchannels $K$ with $d = 500$ kbps. From this figure, we can observe that the number of scheduled IoT devices grows as the number of licensed subchannels and the proposed IoT-U scheme outperforms IoT-L scheme. This implies that utilizing the unlicensed spectrum to support M2M communications can effectively mitigate the interference to the cellular links, and thus improving the system performance. In addition, compared to the DASPA and MASPA schemes, the IASPA algorithm in IoT-U scheme can schedule more IoT devices.

Fig.~\ref{Figure2} shows the energy consumption v.s. the number of licensed subchannels $K$ with $d = 500$ kbps. It is observed that the energy consumption increases as the number of licensed subchannels because more IoT devices are scheduled. Combining these two figures, we can infer that the TC network is the most energy efficient, that is, costing the least energy for one scheduled IoT device. This implies that the cluster based IoT networks can schedule more IoT devices at the cost of more energy for the CU relay. Moreover, we can find out that the energy consumption of an IoT device in the IoT-U scheme is lower than that in the DASPA and MASPA schemes because the IASPA algorithm in the IoT-U scheme can utilize the resource more efficiently.

\begin{figure}
\centering
\subfigure[]{
\label{Figure3}
\includegraphics[height=2.3in]{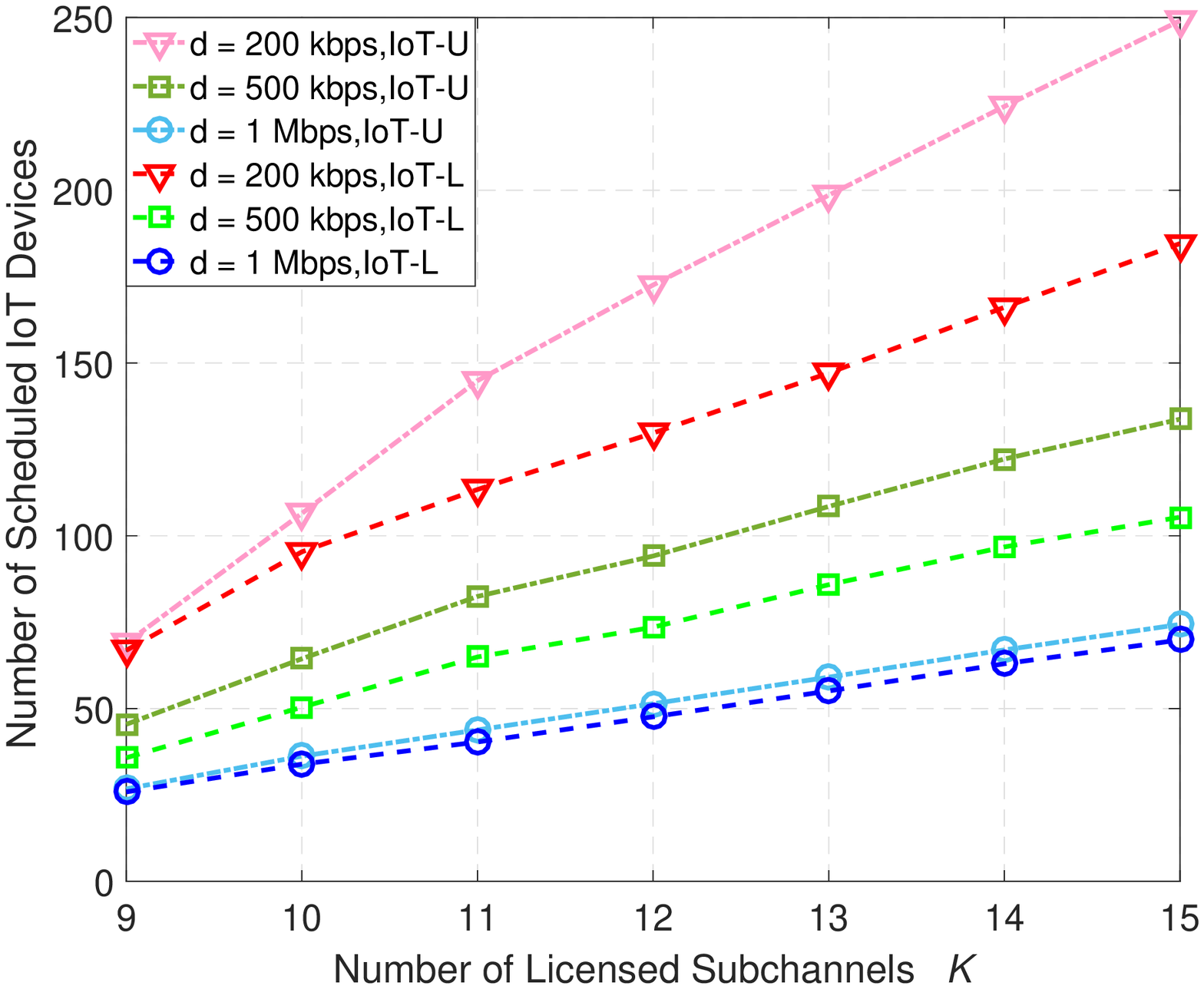}}
\subfigure[]{
\label{Figure4}
\includegraphics[height=2.3in]{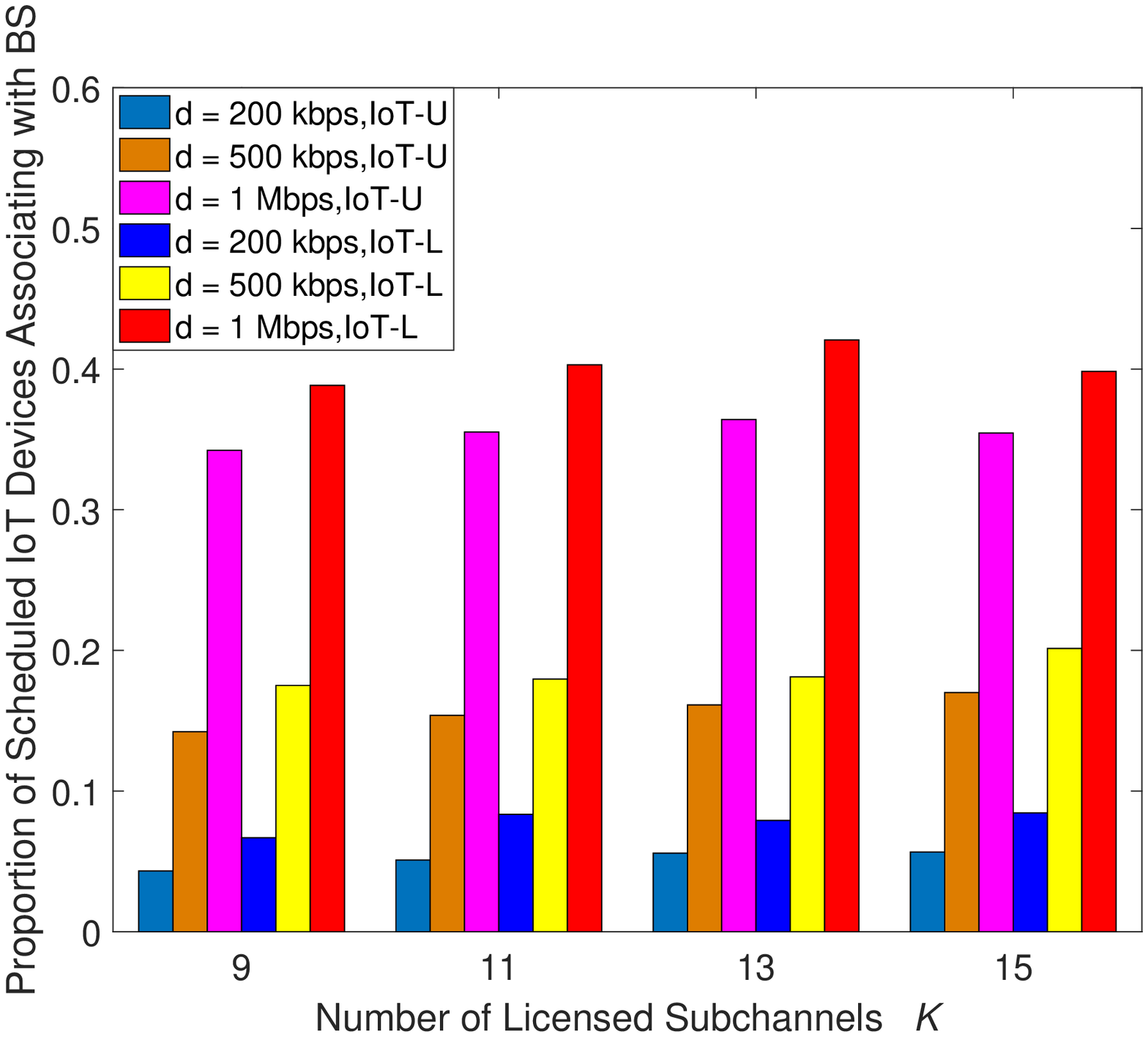}}
\caption{(a) Number of scheduled IoT devices v.s. number of licensed subchannels; (b) Proportion of scheduled IoT devices associating with BS v.s. number of licensed subchannels.}\label{Figure34}
\vspace{-5mm}	
\end{figure}

Fig.~\ref{Figure34} illustrates how the data traffic influence the system performance with respect to the number of scheduled IoT devices and the IoT device association strategies. In Fig.~\ref{Figure3}, the number of scheduled IoT devices is evaluated for both IoT-L and IoT-U schemes with different data traffic of a IoT device. When the data traffic is low, a CU can aggregate the data traffic from multiple IoT devices. However, due to the subchannel sharing, the interference may be significant when a relatively large number of M2M links coexist on the same licensed subchannel. IoT-U scheme offloads M2M communications into the unlicensed band and alleviates the interference. Therefore, more IoT devices are scheduled. As the data traffic grows, the number of M2M links in a cluster decreases and the benefit from utilizing unlicensed subchannels for M2M communications also decreases. In addition, when the data traffic is low, i.e., $d = 200$ kbps, we can find out that the growth rate for IoT-U scheme with $K < 6$ is higher than that with $K \geq 6$. When $K < 6$, the number of scheduled IoT devices is bounded by the number of subchannels and it is bounded by the data rate of cellular links when $K \geq 6$, which are in accordance with Proposition \ref{proposition} in Section \ref{Association}. 

Fig.~\ref{Figure4} further evaluates how different data traffic influences the association strategy of IoT devices. We can easily observe that the proportion of scheduled IoT devices increases when the data traffic of each IoT devices becomes larger. Note that the data rate for the cellular link becomes the bottleneck when the data traffic for a IoT devices is high. Therefore, more IoT devices are associated with the BS using the orthogonal subchannels to avoid interference to other cellular links. This is consistent with Remark \ref{remark} in Section \ref{Association}. Besides, we can also observe that the proportion of scheduled IoT devices associating with the BS obtained by IoT-U scheme is less than that obtained by IoT-L scheme. Since M2M links may bring severe interference to the cellular links in IoT-L scheme, the IoT devices located far from the CUs need to associate with the BS. However, in IoT-U scheme, the interference from M2M links can be mitigated by allocating the transmit power to the unlicensed subchannels, and thus, a cluster can support more IoT devices than that in IoT-L scheme.

\begin{figure}[!t]
\centering
\includegraphics[height=2.8in]{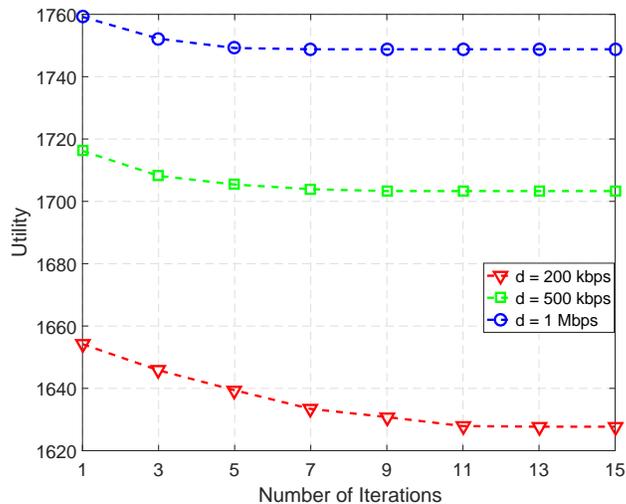}
\caption{Utility v.s. number of iterations with $K = 7$.}\label{Figure5}
\vspace{-5mm}	
\end{figure}

Fig.~\ref{Figure5} shows the convergence of the proposed IASPA algorithm with 12 licensed subchannels for different values of $d$. It can be observed that 11 iterations are needed when $d = 200$ kbps while it requires 7 iterations when $d = 1$ Mbps, which implies that higher data rate for an IoT device requires more iterations. When the data rate for one IoT device is high, the remaining transmit power of CU may not support the data from another IoT device. Therefore, the algorithm with a  higher data rate requires more iterations. 

\section{Conclusions}
\label{sec:conclusion}
In this paper, we proposed the IoT-U scheme for the cluster based IoT network where the unlicensed spectrum can be utilized for the M2M communications in clusters. We first designed a duty cycle based mechanism for fair unlicensed spectrum sharing, and then formulated an optimization problem to maximize the total weighted number of scheduled IoT devices with the minimum transmission energy consumption.  Simulation results show that the proposed IoT-U scheme can support more IoT devices than the IoT-L one in one cycle. From analysis, we can obtain two important conclusions. First, since the IoT-U scheme offloads M2M communication into the unlicensed band and alleviates the co-channel interference, the CU can aggregate the data from more IoT devices in the unlicensed scheme. Second, the IoT devices are intended to associate with CUs when the data traffic is low. But when the data traffic grows, more IoT devices are inclined to associate with the BS to produce less interference. 

\begin{appendices}
	
\section{Proof of Theorem \ref{equilvant}}\label{proof_equilvant}
Let $(\bm{A}^{*},\bm{B}^{*},\bm{C}^{*},\bm{F}^{*},\bm{P}^{*})$ be the optimal solution to problem (P3). We first prove that  $(\bm{A}^{*},\bm{B}^{*},\\\bm{C}^{*},\bm{F}^{*})$ yields the maximum value of the objective in problem (P1), and then prove that $\bm{P}^{*}$ yields the minimum total transmission power in problem (P2). 

To prove that $(\bm{A}^{*},\bm{B}^{*},\bm{C}^{*},\bm{F}^{*})$ yields the maximum value of the objective in problem (P1) is equivalent to prove that $\sum\limits_{m \in \mathcal{M}}w_m\sum\limits_{t \in \mathcal{T}}c^{*}_{m,t} = \sum\limits_{m \in \mathcal{M}}w_m\sum\limits_{t \in \mathcal{T}}\tilde{c}_{m,t}$, where $(\bm{\tilde{A}},\bm{\tilde{B}},\bm{\tilde{C}},\bm{\tilde{F}},\bm{\tilde{P}})$ is an optimal solution of problem (P1). Note that the constraints in problem (P3) is the same as that in problem (P1), and thus, the optimal solution of problem (P3) is also feasible to problem (P1). Suppose that $(\bm{\tilde{A}},\bm{\tilde{B}},\bm{\tilde{C}},\bm{\tilde{F}})$ yields a higher value of the objective than $(\bm{A}^{*},\bm{B}^{*},\bm{C}^{*},\bm{F}^{*},)$, i.e.,
\begin{equation}\label{eq1_proof}
\sum\limits_{m \in \mathcal{M}}w_m\sum\limits_{t \in \mathcal{T}}\tilde{c}_{m,t} > \sum\limits_{m \in \mathcal{M}}w_m\sum\limits_{t \in \mathcal{T}}c^{*}_{m,t}.
\end{equation}
By multiplying $\rho$ to the both sides of (\ref{eq1_proof}), we can obtain
\begin{equation}\label{eq2_proof}
\sum\limits_{m \in \mathcal{M}}\lambda_m\sum\limits_{t \in \mathcal{T}}\tilde{c}_{m,t} > \sum\limits_{m \in \mathcal{M}}\lambda_m\sum\limits_{t \in \mathcal{T}}c^{*}_{m,t}.
\end{equation}
Since $c^{*}_{m,t}$, $\tilde{c}_{m,t}$, and $\lambda_m$ are all positive integers, (\ref{eq2_proof}) can be tightened to
\begin{equation}\label{eq3_proof}
\sum\limits_{m \in \mathcal{M}}\lambda_m\sum\limits_{t \in \mathcal{T}}\tilde{c}_{m,t} - 1 \geq \sum\limits_{m \in \mathcal{M}}\lambda_m\sum\limits_{t \in \mathcal{T}}c^{*}_{m,t}.
\end{equation}  

Note that $0 < \epsilon < \frac{1}{MP^{iot} + NP^{c} + 1}$. Therefore, we can obtain the following inequality:
\begin{equation}\label{eq4_proof}
\sum\limits_{m \in \mathcal{M}} \sum\limits_{t \in \mathcal{T}} \sum\limits_{k \in \mathcal{K} \cup \mathcal{K}^u} \sum\limits_{n \in \mathcal{Q} \cup \{0\}} \epsilon p_{m,n}^{k,t} +\sum\limits_{n \in \mathcal{N}} \sum\limits_{t \in \mathcal{T}} \sum\limits_{k \in \mathcal{K}} \epsilon p_{n}^{k,t} < \epsilon (MP^{iot} + NP^{c}) < 1-\epsilon.
\end{equation}  
 Thus, together with (\ref{eq3_proof}), the objective function values corresponding to $(\bm{A}^{*},\bm{B}^{*},\bm{C}^{*},\bm{F}^{*},\bm{P}^{*})$ and $(\bm{\tilde{A}},\bm{\tilde{B}},\bm{\tilde{C}},\bm{\tilde{F}},\bm{\tilde{P}})$ satisfy
 \begin{equation}\label{eq5_proof}
 \begin{array}{ll}
 & \sum\limits_{m \in \mathcal{M}} \sum\limits_{t \in \mathcal{T}} \sum\limits_{k \in \mathcal{K} \cup \mathcal{K}^u} \sum\limits_{n \in \mathcal{Q} \cup \{0\}} \epsilon \tilde{p}_{m,n}^{k,t} +\sum\limits_{n \in \mathcal{N}} \sum\limits_{t \in \mathcal{T}} \sum\limits_{k \in \mathcal{K}} \epsilon \tilde{p}_{n}^{k,t} + (1-\epsilon) \sum\limits_{m \in \mathcal{M}}\lambda_m\sum\limits_{t \in \mathcal{T}}(1 - \tilde{c}_{m,t})\\
 \leq &\sum\limits_{m \in \mathcal{M}} \sum\limits_{t \in \mathcal{T}} \sum\limits_{k \in \mathcal{K} \cup \mathcal{K}^u} \sum\limits_{n \in \mathcal{Q} \cup \{0\}} \epsilon \tilde{p}_{m,n}^{k,t} +\sum\limits_{n \in \mathcal{N}} \sum\limits_{t \in \mathcal{T}} \sum\limits_{k \in \mathcal{K}} \epsilon \tilde{p}_{n}^{k,t} + (1-\epsilon) \sum\limits_{m \in \mathcal{M}}\lambda_m\sum\limits_{t \in \mathcal{T}}(1 - c^{*}_{m,t}) - (1 - \epsilon)\\
  < &(1-\epsilon) \sum\limits_{m \in \mathcal{M}}\lambda_m\sum\limits_{t \in \mathcal{T}}(1 - c^{*}_{m,t}).
 \end{array}
 \end{equation} 
 We can observe that $(\bm{\tilde{A}},\bm{\tilde{B}},\bm{\tilde{C}},\bm{\tilde{F}},\bm{\tilde{P}})$ yields a smaller objective function value than $(\bm{A}^{*},\bm{B}^{*},\bm{C}^{*},\\\bm{F}^{*},\bm{P}^{*})$ does. Since $(\bm{\tilde{A}},\bm{\tilde{B}},\bm{\tilde{C}},\bm{F}^{*},\bm{\tilde{P}})$ is also feasible to problem (P3), (\ref{eq5_proof}) violates the assumption that $(\bm{A}^{*},\bm{B}^{*},\bm{C}^{*},\bm{F}^{*},\bm{P}^{*})$ is the optimal solution of problem (P3). Therefore, we have 
 \begin{equation}\label{eq6_proof}
 \sum\limits_{m \in \mathcal{M}}w_m\sum\limits_{t \in \mathcal{T}}\tilde{c}_{m,t} = \sum\limits_{m \in \mathcal{M}}w_m\sum\limits_{t \in \mathcal{T}}c^{*}_{m,t}.
 \end{equation}
 
 Following the same approach, we can also prove that  $\bm{P}^{*}$ yields the minimum total transmission power in problem (P2).  
 
On the other hand, let $(\bm{A}^{*},\bm{B}^{*},\bm{C}^{*},\bm{F}^{*},\bm{P}^{*})$ be the optimal solution through solving problems (P1) and (P2) sequentially. We also prove that $(\bm{A}^{*},\bm{B}^{*},\bm{C}^{*},\bm{F}^{*},\bm{P}^{*})$ is the optimal solution to problem (P3). Define $(\bm{\tilde{A}},\bm{\tilde{B}},\bm{\tilde{C}},\bm{\tilde{F}},\bm{\tilde{P}})$ as the optimal solution of problem (P3). To prove that $(\bm{A}^{*},\bm{B}^{*},\bm{C}^{*},\bm{F}^{*},\bm{P}^{*})$ is the optimal solution to problem (P3) is equivalent to prove that $(\bm{A}^{*},\bm{B}^{*},\bm{C}^{*},\bm{F}^{*},\bm{P}^{*})$ can yield to the same value of $U$ as $(\bm{\tilde{A}},\bm{\tilde{B}},\bm{\tilde{C}},\bm{\tilde{F}},\bm{\tilde{P}})$ does.
 
 Note that the feasible set of problem (P2) is the optimal solution set of problem (P1). Therefore, $(\bm{A}^{*},\bm{B}^{*},\bm{C}^{*},\bm{F}^{*},\bm{P}^{*})$ yields the maximum value of the objective in problem (P1). Since the constraints in problem (P1) are the same as that in problem (P3), the optimal solution of problem (P1) is also feasible to problem (P3). Suppose that  $(\bm{\tilde{A}},\bm{\tilde{B}},\bm{\tilde{C}},\bm{\tilde{F}},\bm{\tilde{P}})$ yields a lower value of the objective than $(\bm{A}^{*},\bm{B}^{*},\bm{C}^{*},\bm{F}^{*},\bm{P}^{*})$, that is,
 \begin{equation}
 \begin{array}{ll}
 &\sum\limits_{m \in \mathcal{M}}\sum\limits_{t \in \mathcal{T}}\sum\limits_{k \in \mathcal{K} \cup \mathcal{K}^u} \sum\limits_{n \in \mathcal{Q} \cup \{0\}}\epsilon\tilde{p}_{m,n}^{k,t} + \sum\limits_{n \in \mathcal{N}} \sum\limits_{t \in \mathcal{T}} \sum \limits_{k \in \mathcal{K}} \epsilon \tilde{p}_n^{k,t} + (1 - \epsilon)\sum\limits_{m \in \mathcal{M}}\lambda_m\sum\limits_{t \in \mathcal{T}}(1 - \tilde{c}_{m,t})\\
 < &\sum\limits_{m \in \mathcal{M}}\sum\limits_{t \in \mathcal{T}}\sum\limits_{k \in \mathcal{K} \cup \mathcal{K}^u} \sum\limits_{n \in \mathcal{Q} \cup \{0\}}\epsilon(p_{m,n}^{k,t})^{*} + \sum\limits_{n \in \mathcal{N}} \sum\limits_{t \in \mathcal{T}} \sum \limits_{k \in \mathcal{K}} \epsilon (p_n^{k,t})^{*} + (1 - \epsilon)\sum\limits_{m \in \mathcal{M}}\lambda_m\sum\limits_{t \in \mathcal{T}}(1 - c^{*}_{m,t}).
 \end{array}
 \end{equation} 

Similarly, $(\bm{\tilde{A}},\bm{\tilde{B}},\bm{\tilde{C}},\bm{\tilde{F}},\bm{\tilde{P}})$ is also a feasible solution of problem (P1). If $\sum\limits_{m \in \mathcal{M}}\lambda_m\sum\limits_{t \in \mathcal{T}}\tilde{c}_{m,t} = \sum\limits_{m \in \mathcal{M}}\lambda_m\sum\limits_{t \in \mathcal{T}}c^{*}_{m,t}$, then $(\bm{A}^{*},\bm{B}^{*},\bm{C}^{*},\bm{F}^{*},\bm{P}^{*})$ is also the optimal solution to problem (P3), because $\bm{P}^{*}$ yields to the minimum total transmission power. Therefore, we have 
\begin{equation}\label{eq7_proof}
\sum\limits_{m \in \mathcal{M}}\lambda_m\sum\limits_{t \in \mathcal{T}}\tilde{c}_{m,t} < \sum\limits_{m \in \mathcal{M}}\lambda_m\sum\limits_{t \in \mathcal{T}}c^{*}_{m,t}.
\end{equation}
Since $c^{*}_{m,t}$, $\tilde{c}_{m,t}$, and $\lambda_m$ are all positive integers, (\ref{eq7_proof}) can be tightened to
\begin{equation}
 (1 - \epsilon)\sum\limits_{m \in \mathcal{M}}\lambda_m\sum\limits_{t \in \mathcal{T}}(1 - \tilde{c}_{m,t}) - (1 - \epsilon) \geq (1 - \epsilon)\sum\limits_{m \in \mathcal{M}}\lambda_m\sum\limits_{t \in \mathcal{T}}(1 - c^{*}_{m,t}).
\end{equation}
Therefore, we have
\begin{equation}
\begin{array}{ll}
&(1 - \epsilon)\sum\limits_{m \in \mathcal{M}}\lambda_m\sum\limits_{t \in \mathcal{T}}(1 - \tilde{c}_{m,t}) - (1 - \epsilon) + \sum\limits_{m \in \mathcal{M}}\sum\limits_{t \in \mathcal{T}}\sum\limits_{k \in \mathcal{K} \cup \mathcal{K}^u} \sum\limits_{n \in \mathcal{Q} \cup \{0\}}\epsilon(p_{m,n}^{k,t})^{*} + \sum\limits_{n \in \mathcal{N}} \sum\limits_{t \in \mathcal{T}} \sum \limits_{k \in \mathcal{K}} \epsilon (p_n^{k,t})^{*}\\
&\geq (1 - \epsilon)\sum\limits_{m \in \mathcal{M}}\lambda_m\sum\limits_{t \in \mathcal{T}}(1 - c^{*}_{m,t}) + \sum\limits_{m \in \mathcal{M}}\sum\limits_{t \in \mathcal{T}}\sum\limits_{k \in \mathcal{K} \cup \mathcal{K}^u} \sum\limits_{n \in \mathcal{Q} \cup \{0\}}\epsilon(p_{m,n}^{k,t})^{*} + \sum\limits_{n \in \mathcal{N}} \sum\limits_{t \in \mathcal{T}} \sum \limits_{k \in \mathcal{K}} \epsilon (p_n^{k,t})^{*}\\
& > \sum\limits_{m \in \mathcal{M}}\sum\limits_{t \in \mathcal{T}}\sum\limits_{k \in \mathcal{K} \cup \mathcal{K}^u} \sum\limits_{n \in \mathcal{Q} \cup \{0\}}\epsilon\tilde{p}_{m,n}^{k,t} + \sum\limits_{n \in \mathcal{N}} \sum\limits_{t \in \mathcal{T}} \sum \limits_{k \in \mathcal{K}} \epsilon \tilde{p}_n^{k,t} + (1 - \epsilon)\sum\limits_{m \in \mathcal{M}}\lambda_m\sum\limits_{t \in \mathcal{T}}(1 - \tilde{c}_{m,t}),
\end{array}
\end{equation}
that is,
\begin{equation}\label{eq8_proof}
\begin{array}{ll}
& \sum\limits_{m \in \mathcal{M}}\sum\limits_{t \in \mathcal{T}}\sum\limits_{k \in \mathcal{K} \cup \mathcal{K}^u} \sum\limits_{n \in \mathcal{Q} \cup \{0\}}\epsilon(p_{m,n}^{k,t})^{*} + \sum\limits_{n \in \mathcal{N}} \sum\limits_{t \in \mathcal{T}} \sum \limits_{k \in \mathcal{K}} \epsilon (p_n^{k,t})^{*}  - (1 - \epsilon)\\
> &\sum\limits_{m \in \mathcal{M}}\sum\limits_{t \in \mathcal{T}}\sum\limits_{k \in \mathcal{K} \cup \mathcal{K}^u} \sum\limits_{n \in \mathcal{Q} \cup \{0\}}\epsilon\tilde{p}_{m,n}^{k,t} + \sum\limits_{n \in \mathcal{N}} \sum\limits_{t \in \mathcal{T}} \sum \limits_{k \in \mathcal{K}} \epsilon \tilde{p}_n^{k,t}  \geq 0.
\end{array}
\end{equation}
Note that $0 < \epsilon < \frac{1}{MP^{iot} + NP^{c} + 1}$. Therefore, we can obtain the following inequality:
\begin{equation}
\sum\limits_{m \in \mathcal{M}} \sum\limits_{t \in \mathcal{T}} \sum\limits_{k \in \mathcal{K} \cup \mathcal{K}^u} \sum\limits_{n \in \mathcal{Q} \cup \{0\}} \epsilon (p_{m,n}^{k,t})^{*} +\sum\limits_{n \in \mathcal{N}} \sum\limits_{t \in \mathcal{T}} \sum\limits_{k \in \mathcal{K}} \epsilon (p_{n}^{k,t})^{*} < \epsilon (MP^{iot} + NP^{c}) < 1-\epsilon,
\end{equation} 
which violates (\ref{eq8_proof}). Therefore, $(\bm{A}^{*},\bm{B}^{*},\bm{C}^{*},\bm{F}^{*},\bm{P}^{*})$ is also the optimal solution to problem (P3).

\end{appendices}


\begin{thebibliography}{40}
	
\bibitem{M-2015}	
Mckinsey Global Institute, ``The Internet of Things: Mapping the Value Beyond the Hype," Jun. 2015.

\bibitem{MKM-2018}
M. Tao, K. Ota, and M. Dong, ``Locating compromised data sources in IoT-enabled smart cities: A great-alternative-region-based approach," \emph{IEEE Trans. Ind. Informat.}, vol.~14, no.~6, pp.~2579-2587, Jun. 2018.


\bibitem{OTMJM-2017}
K. Ota, T. Kumrai, M. Dong, J. Kishigami, and M. Guo, ``Smart infrastructure design for smart cities," \emph{IT Professional}, vol.~19, no.~5, pp.~42-49, Sep.~2017.

\bibitem{SPS-2017}
S.~Feng, P.~Setoodeh, and S.~Haykin, ``Smart Home: Cognitive Interactive People-Centric Internet of Things," \emph{IEEE Commun. Mag.}, vol.~55, no.~2, pp.~34-39, Feb.~2017.

\bibitem{UBM-2017}
U. Satija, B. Ramkumar, and M. S. Manikandan, ``Real-Time Signal Quality-Aware ECG Telemetry System for IoT-Based Health Care Monitoring," \emph{IEEE Internet Things J.}, vol.~4, no.~3, pp.~815-823, Jun.~2017.

\bibitem{CINKK-2017}
C.~Brewster, I.~Roussaki, N.~Kalatzis, K.~Doolin, and K.~Ellis, ``IoT in Agriculture: Designing a Europe-Wide Large-Scale Pilot," \emph{IEEE Commun. Mag.}, vol.~55, no.~9, pp.~26-33, Sep.~2017.

\bibitem{AXOAEMYTAD-2017}
A.~Hoglund, X.~Lin, O.~Liberg, A.~Behravan, E.~A.~Yavuz, M.~V.~D.~Zee, Y.~Sui, T.~Tirronen, A.~Ratilainen, and D.~Eriksson, ``Overview of 3GPP Release 14 Enhanced NB-IoT," \emph{IEEE Network}, vol.~31, no.~6, pp.~16-22, Nov.~2017.




\bibitem{HYL-2017}
H.~Zhang, Y.~Liao, and L.~Song, ``D2D-U: Device-to-Device Communications in Unlicensed Bands for 5G System," \emph{IEEE Trans. Wireless Commun.}, vol.~16, no.~6, pp.~3507-3519, Jun.~2017.

\bibitem{PBHKL-2018}
P.~Wang, B.~Di, H.~Zhang, K.~Bian, and L.~Song, ``Cellular V2X in Unlicensed Spectrum: Harmonious Coexistence with VANET in 5G systems," \emph{IEEE Trans. Wireless Commun.}, vol.~17, no.~8, pp.~5212-5224, Aug.~2018.

\bibitem{RMLZXL-2015}
R.~Zhang, M.~Wang, L.~X.~Cai, Z.~Zeng, X.~S.~Shen, and L.~Xie, ``LTE-Unlicensed: The Future of Spectrum Aggregation for Cellular Networks," \emph{IEEE Wireless Commun.}, vol.~22, no.~3, pp.~150-159, Jun.~2015.

\bibitem{ZMKGL-2016}
Z. Zhou, M. Dong, K. Ota, G. Wang, and L. T. Yang, ``Energy-Efficient Resource Allocation for D2D Communications Underlaying Cloud-RAN-based LTE-A Networks," \emph{IEEE J. Internet Things},  vol. 3, no. 3, pp. 428-438, Jun. 2016.

\bibitem{FQ-2016}
F. Ramparany and Q. H. Cao, ``A Semantic Approach to IoT Data Aggregation and Interpretation Applied to Home Automation," in \emph{Proc. IOTA}, Pune, India, Jan. 2016.

\bibitem{LoRa-2015}
LoRa Alliance, ``A Technical Overview of LoRa and LoRaWAN", Nov. 2015.

\bibitem{MLAM-2016}
M.~Centenaro, L.~Vangelista, A.~Zanella, and M.~Zorzi, ``Long-Range Communications in Unlicensed Bands: The Rising Stars in the IoT and Smart City Scenarios," \emph{IEEE Wireless Commun.}, vol.~23, no.~5, pp.~60-67, Oct.~2016.

\bibitem{TVOMAMSRN-2013}
T.~Nihtil\"a, V.~Tykhomyrov, O.~Alanen, M.~A.~Uusitalo, A.~Sorri, M.~Moisio, S.~Iraji, R.~Ratasuk, and N.~Mangalvedhe, ``System Performance of LTE and IEEE 802.11 Coexisting on a Shared Frequency Band," in \emph{Proc. IEEE WCNC}, Shanghai, China, Apr.~2013.

\bibitem{Qualcomm-2014}
Qualcomm Research, ``LTE in Unlicensed Spectrum: Harmonious Coexistence with Wi-Fi," Jun. 2014.

\bibitem{ED-1966}
E.~L.~Lawler and D.~E.~Wood, ``Branch-and-Bound Methods: A Survey," \emph{Operations Research}, vol.~14, no.~4, pp.~699-719, Aug.~1966.

\bibitem{JJ-2009}
J.~Papandriopoulos and J.~S.~Evans, ``SCALE: A Low-complexity Distributed Protocol for Spectrum Balancing in Multiuser DSL Networks," \emph{IEEE Trans. Inf. Theory}, vol.~55, no.~8, pp.~3711-3724, Aug.~2009.

\bibitem{SL-2004}
S.~Boyd and L.~Vandenberghe, \emph{Convex Optimization}. Cambridge, U.K.: Cambridge University Press, 2004.

\bibitem{MBWAF-2015}
M.~Darabi, B.~Maham, W.~Saad, A.~Mehbodniya, and F.~Adachi, ``Joint Machine-Type Device Selection and Power Allocation for Buffer-Aided Cognitive M2M Communication," in \emph{Proc. IEEE PIMRC}, Hong Kong, China, Sep. 2015.

\bibitem{MABWH-2018}
M.~Darabi, A.~Montazeri, B.~Maham, W.~Saad, and H.~Zarrabi, ``Packet Size Adjustment for Minimizing the Average Delay in Buffer-Aided Cognitive Machine-to-Machine Networks," \emph{Elsevier J. Comput. Elect. Eng.}, vol.~68, pp.~298–309, May 2018.

\bibitem{XYSH-2014}
X. Kang, Y.-K. Chia, S. Sun, and H. F. Chong, ``Mobile Data Offloading Through A Third-Party Wi-Fi Access Point: An Operator's Perspective," \emph{IEEE Trans. Wireless Commun.}, vol. 13, no. 10, pp. 5340-5351, Oct. 2014.

\bibitem{KJYIS-2013}
K. Lee, J. Lee, Y. Yi, I. Rhee, and S. Chong, ``Mobile Data Offloading: How Much Can Wi-Fi Deliver?" \emph{IEEE/ACM Trans. Netw.}, vol. 21, no. 2, pp. 536-550, Apr. 2013.

\bibitem{Nokia-2014}
Nokia, ``Nokia LTE for Unlicensed Spectrum," Jun. 2014.

\bibitem{HXWS-2015}
H. Zhang, X. Chu, W. Guo, and S. Wang, ``Coexistence of Wi-Fi and Heterogeneous Small Cell Networks Sharing Unlicensed Spectrum," \emph{IEEE Commun. Mag.}, vol.~53, no.~3, pp.~158-164, Mar.~2015.

\bibitem{MSMS-2017}
M. Ali, S. Qaisar, M. Naeem, and S. Mumtaz, ``Joint User Association and Power Allocation for Licensed and Unlicensed Spectrum in 5G Networks," in \emph{Proc. IEEE GLOBECOM}, Singapore, Singapore, Dec. 2017.

\bibitem{BJYJ-2017}
B. Chen, J. Chen, Y. Gao, and J. Zhang, ``Coexistence of LTE-LAA and Wi-Fi on 5GHz with Corresponding Deployment Scenarios: A Survey," \emph{IEEE Commun.  Surveys \& Tutorials}, vol. 19, no. 1, pp. 7-32, First Quarter 2017.

\bibitem{JWSS-2018}
J. Yi, W. Sun, S. Park, and S. Choi, ``Performance Analysis of LTE-LAA Network," \emph{IEEE Commun. Lett.}, vol. 22, no. 6, pp. 1236-1239, Jun. 2018. 




\bibitem{3GPP-2017}
3GPP TR 36.746, ``Study on Further Enhancements to LTE Device to Device (D2D), User Equipment (UE) to Network Relays for Internet of Things (IoT) and Wearables," Apr. 2017. 

\bibitem{BLY-2016}
B.~Di, L.~Song, and Y.~Li, ``Sub-Channel Assignment, Power Allocation, and User Scheduling for Non-Orthogonal Multiple Access Networks," \emph{IEEE Trans. Wireless Commun.}, vol.~15, no.~11, pp.~7686-7698, Nov.~2016.

\bibitem{LYYJ-2013}
L.~P.~Qian, Y.~J.~Zhang, Y.~Wu, and J.~Chen, ``Joint Base Station Association and Power Control via Benders' Decomposition," \emph{IEEE Trans. Wireless Commun.}, vol.~12, no.~4, pp.~1651-1665, Apr.~2013.

\bibitem{MSAJ-2007}
M.~Chiang, S.~H.~Low, A.~R.~Calderbank, and J.~C.~Doyle, ``Layering as Optimization Decomposition: A Mathematical Theory of Network Architectures," \emph{Proc. IEEE}, vol.~95, no.~1, pp.~255-312, Jan.~2007.

\bibitem{DX-2006}
D.~Li and X.~Sun, \emph{Nonlinear Integer Programming}. Boston, MA: Springer Verlag, 2006.

\bibitem{AGJ-2014}
A.~Alvarado, G.~Scutari, and J.-S.~Pang, ``A New Decomposition Method for Multiuser
DC-Programming and Its Applications," \emph{IEEE Trans. Signal Process.}, vol.~62, no.~11, pp.~2984-2998, Jun.~2014.

\bibitem{S-2013}
S.~Boyd, ``Subgradient Methods", [Online]. Available: http://www.stanford.edu/class/ee364b/lectures/subgrad\underline{\hspace{0.5em}}method\underline{\hspace{0.5em}}notes.pdf.

\bibitem{A-1998}
A. Schrijver, \emph{Theory of Linear and Integer Programming}, John Wiley \& Sons, New Jersey, USA, 1998.


\bibitem{TNWLDTC-2017}
T. LeAnh, N. H. Tran, W. Saad, L. B. Le, D. Niyato, T. M. Ho, and C. S. Hong, ``Matching Theory for Distributed User Association and Resource Allocation in Cognitive Femtocell Networks," \emph{IEEE Trans. Veh. Technol.}, vol. 66, no. 9, pp. 8413-8428, Sep. 2017.

\bibitem{3GPP-2013}
3GPP TR 36.888, ``Study on Provision of Low-cost Machine-Type Communications (MTC) User Equipments (UEs) based on LTE (Release 12)," Jun.~2013.

\end{thebibliography}
\end{document}